# Realization of the Axion Insulator State in Quantum Anomalous Hall Sandwich Heterostructures

Di Xiao[1]*, Jue Jiang[1]*, Jae-Ho Shin[1], Wenbo Wang[2], Fei Wang[1], Yi-Fan Zhao[1], Chaoxing Liu[1], Weida Wu[2], Moses H. W. Chan[1], Nitin Samarth[1] and Cui-Zu Chang[1]

[1]Department of Physics, Pennsylvania State University, University Park, PA16802, USA

[2]Department of Physics and Astronomy, Rutgers University, Piscataway, NJ 08854, USA

*These authors contributed equally to this work.

Corresponding authors: mhc2@psu.edu (M.H.W.C.); nxs16@psu.edu (N.S.); cxc955@psu.edu (C.Z.C.)

The 'magnetoelectric effect' arises from the coupling between magnetic and electric properties in materials. The $Z_2$ invariant of topological insulators (TIs) leads to a quantized version of this phenomenon, known as the topological magnetoelectric (TME) effect. This effect can be realized in a new topological phase called an 'axion insulator' whose surface states are all gapped but the interior still obeys time reversal symmetry. We demonstrate such a phase using electrical transport measurements in a quantum anomalous Hall (QAH) sandwich heterostructure, in which two compositionally different magnetic TI layers are separated by an undoped TI layer. Magnetic force microscopy images of the same sample reveal sequential magnetization reversals of the top and bottom layers at different coercive fields, a consequence of the weak interlayer exchange coupling due to the spacer. When the magnetization is antiparallel, both the Hall resistance and Hall conductance show zero plateaus, accompanied by a large longitudinal resistance and vanishing longitudinal conductance, indicating the realization of an **axion insulator state. Our findings thus show evidences for a phase of matter distinct from the established QAH state and provide a promising platform for the realization of the TME effect.**

Three-dimensional (3D) topological insulators (TIs) are unusual quantum materials that host conducting helical Dirac states on their surfaces, which are protected by time reversal symmetry (TRS), but are electrically insulating in the bulk [1,2]. TI is distinct from a trivial insulator by its unique electromagnetic response, described by the so-called $\theta$ term shown below in addition to the ordinary Maxwell terms [3-6].

$$S_\theta = \frac{\theta}{2\pi} \frac{e^2}{h} \int d^3 x dt \vec{E} \cdot \vec{B}$$

Here $\vec{E}$ and $\vec{B}$ are the conventional electric and magnetic fields inside an insulator, $e$ is electron charge, and $\theta$ is the dimensionless pseudo-scalar parameter describing the insulator. For a trivial insulator, $\theta=0$, while for a TI, $\theta=\pi$. When TRS is preserved, $\theta$ is either 0 or π, reflecting its topological nature. This $\theta$ term is related to the axion electrodynamics in particle physics [7]. Since the $\vec{E} \cdot \vec{B}$ term can be rewritten as a total derivative, its effect manifests on the surface states. A half-integer quantum Hall effect on the TI surface occurs once the surface Dirac Fermions acquire a mass, *i.e.* the surface state is gapped by magnetism. Such half-integer quantum Hall effect on TI surface can lead to a variety of exotic phenomena such as the quantum anomalous Hall (QAH) effect [3,8-15], the quantized magneto-optical effect [3,16,17], the topological magnetoelectric (TME) effect [3-6,18], and the image magnetic monopole [19]. The QAH and quantized magneto-optical effects have been experimentally demonstrated in pure or magnetic TI films [10-13,20-22]. The TME effect refers to the quantized response of electric polarization to applied magnetic fields and vice versa. The realization of the TME effect requires the following three conditions: (i) the TI film should be in the 3D regime; (ii) all the surfaces are gapped with the chemical potential lying within the gaps; (iii) the interior of the TI maintains TRS or inversion symmetry to maintain $\theta=\pi$ in the bulk. A material system allowing for the realization of TME effect is known as an axion insulator [3-5].

Recently, two papers reported the possible realization of the axion insulator. Mogi *et al* [23] fabricated asymmetric (Bi,Sb)$_2$Te$_3$ multilayer heterostructures with Cr modulation doping. The observation of a zero Hall conductance ($\sigma_{xy}$) plateau was interpreted as evidence of an axion insulator due to antiparallel magnetization alignment of the top and bottom Cr-doped (Bi,Sb)$_2$Te$_3$ layers. Subsequent magnetic domain imaging measurements on the same sample, however, failed to find evidence of antiparallel magnetization alignment at any external magnetic field ($\mu_0H$) [24]. It is likely that the reported zero $\sigma_{xy}$ plateau is not a result of antiparallel magnetization alignment but an artifact due to the conversion of the measured longitudinal resistance ($\rho_{xx}$) and Hall resistance ($\rho_{yx}$) into $\sigma_{xy}$, *i. e.* $\sigma_{xy} = \dfrac{\rho_{yx}}{\rho_{xx}^2 + \rho_{yx}^2}$. Nearly zero $\sigma_{xy}$ plateau can arise because of the large $\rho_{xx}$ (~40$h/e^2$) at the coercive field ($H_c$), as shown in Fig. S3h of Ref. [23]. If the top and bottom Cr-doped TI layers indeed possess antiparallel magnetization alignment in the Cr modulation doped QAH samples, then a zero $\rho_{yx}$ plateau should be present, but this is absent in Ref. [23]. Grauer *et al* [25] also claimed the observation of an 'axion insulator' in a uniformly doped QAH sample in the 3D regime. However, the presence of the 1D chiral edge mode of the 3D QAH state implies that side surfaces are not gapped, thus violating the condition (ii) noted above. In addition, the magnetic moments are evenly distributed in the uniformly doped QAH samples, so the interior of the uniformly doped samples no longer maintains TRS, violating condition (iii). We therefore suggest that there is as yet no clear experimental evidence of an axion insulator that satisfies all the necessary criteria as outlined above.

A 3D FM-TI-FM QAH sandwich heterostructure has been proposed for realizing the axion insulator state in antiparallel magnetization alignment [3-5]. It has been shown that magnetically doped TI layers can be epitaxially grown on an undoped TI layer and vice versa to form the required FM-TI-FM sandwich heterostructures [23,26,27]. If the top and bottom $(Bi,Sb)_2Te_3$ layers are doped with two different magnetic ions, specifically, Cr- and V-, and the interlayer coupling exchange field ($H_E$) is substantially smaller than the difference $\Delta H_c$ between the $H_c$s of the two magnetic layers, an antiparallel magnetization alignment may appear when the external $H$ lies between two $H_c$s. This is the case since the $H_c$s of the Cr- and V-doped QAH films [10,13] are respectively ~0.15T and ~1T, as shown in Fig. 1a.

In this *Letter*, we report observations that satisfy the three necessary conditions outlined above for an axion insulator state in V-doped $(Bi,Sb)_2Te_3/(Bi,Sb)_2Te_3$/Cr-doped $(Bi,Sb)_2Te_3$ sandwich heterostructures. Electrical transport and magnetic force microscopy (MFM) measurements show zero $\sigma_{xy}$ and also $\rho_{yx}$ plateaus in the antiparallel magnetization alignment configuration of the top and bottom magnetic TI layers, demonstrating the realization of an axion insulator state. The axion insulator state appears when the $H$ lies between two $H_c$s of the top and bottom magnetic layers.

The V-doped $(Bi,Sb)_2Te_3/(Bi,Sb)_2Te_3$/Cr-doped $(Bi,Sb)_2Te_3$ QAH sandwich heterostructures are grown on $SrTiO_3(111)$ substrate in a molecular beam epitaxy (MBE) chamber. As shown in Fig. 1b, the bottom magnetic layer is 3 quintuple layers (QL) Cr-doped (Bi,Sb)$_2$Te$_3$, and the top magnetic layer is 3QL V-doped (Bi,Sb)$_2$Te$_3$. The undoped (Bi,Sb)$_2$Te$_3$ films separating two magnetic layers have thicknesses of 4, 5 and 6QL. These relatively thick spacer films are chosen to reduce $H_E$ and prevent the hybridization of the top and bottom surface states [25,28-30]. In our experiment, the three different thicknesses of the TI spacer show very similar results, suggesting that a spacer of 4QL is already sufficient in weakening the $H_E$ (see detailed comparisons in Supporting Materials). In this *Letter*, we focus on two sandwich heterostructures with a spacer of 5 QL (referred to as 3-5-3 SH1 and SH2) from the same sample wafer. The samples are scratched into Hall bar geometry (~1.0×0.5mm). A Physical Property Measurement System (PPMS) cryostat is used to initially screen the samples, the principal transport results shown here are then carried out in a dilution refrigerator (Leiden Cryogenics, 10mK, 9T). The MFM experiments are carried out in a homemade cryogenic atomic force microscope with *in-situ* transport measurements.

By carefully tuning the Bi/Sb ratio in each layer of the 3-5-3 sandwich heterostructure, we locate the chemical potential close to its charge neutral point ($V_g^0$). The high dielectric constant of the SrTiO$_3$ substrate at low temperature allows us to fine tune by electric gating the chemical potential within the exchange gap on the top and bottom surfaces. When the magnetization of the top V-doped (Bi,Sb)$_2$Te$_3$ layer is parallel to that of the bottom Cr-doped (Bi,Sb)$_2$Te$_3$ layer, the 3-5-3 sandwich heterostructure exhibits a QAH state with a 1D chiral edge mode flowing along the edge of the sandwich sample, as shown in Fig. 1c. Since $H_c$ of the Cr-doped (Bi,Sb)$_2$Te$_3$ layer is much smaller than that of the V-doped (Bi,Sb)$_2$Te$_3$ layer (Fig. 1a), sweeping the external $\mu_0H$ first reverses the magnetization of the Cr-doped (Bi,Sb)$_2$Te$_3$ layer. Due to the weak $H_E$ between the top and bottom magnetic layers, the V-doped (Bi,Sb)$_2$Te$_3$ layer will remain in its initial magnetization state until the external $H$ reaches $-H_{c1}$. Therefore, when $-H_{c1}<H<-H_{c2}$, the magnetizations of two magnetic layers are in antiparallel alignment. In this case, the top and bottom surfaces can be considered to be in half-integer 'QAH states' with $\sigma^t_{xy}=0.5e^2/h$ and $\sigma^b_{xy}=-0.5e^2/h$, yielding zero $\sigma_{xy}$ plateau and eliminating the 1D chiral edge state. Provided the quantum confinement effect gaps the side surface states, the sandwich structure in antiparallel magnetization alignment regions will satisfy all the conditions for the axion insulator state as illustrated in Fig. 1d.

Figures 2a and 2b show the $\mu_0H$ dependence of the longitudinal conductance ($\sigma_{xx}$) and $\sigma_{xy}$ of the 3-5-3 SH1 measured at $T$=30mK and $V_g=V_g^0$ =+18V. The red and blue curves represent downward and upward $\mu_0H$ sweeps, respectively. At $\pm H_c$s ($\pm H_{c1}$ for the top V-doped TI layer and $\pm H_{c2}$ for the bottom Cr-doped TI layer), we observe sharp peaks of $\sigma_{xx}$ (Fig. 2a). Correspondingly, the $\mu_0H$ dependence of $\sigma_{xy}$ shows a two-step transition from $e^2/h$ to 0 and then $-e^2/h$ or from $-e^2/h$ to 0 and then $e^2/h$. Zero $\sigma_{xy}$ plateaus are observed in the intermediate field ranges $-H_{c1}<H<-H_{c2}$ and $H_{c2}<H<H_{c1}$. We noted that zero $\sigma_{xy}$ plateau has been observed in uniformly doped QAH samples, where it is attributed to large $\rho_{xx}$ resulting from scattering due to the random magnetic domains at $H_c$ [31-33]. In contrast, the zero $\sigma_{xy}$ plateau observed here, as we will show below, results from the cancellation of the top and bottom surface conduction in the antiparallel magnetization alignment, consistent with an axion insulator. In order to demonstrate the antiparallel magnetization alignment in the zero $\sigma_{xy}$ plateau regions, we carried out MFM measurements.

Figures 3a to 3j show *in-situ* $\rho_{yx}$ and MFM images at various $\mu_0H$s on the 3-5-3 SH2 at $T=5.3K$ and $V_g=0V$. The $\rho_{yx}$ shows a two-step transition, similar to the zero $\rho_{yx}$ plateau at $T=30mK$ and $V_g=V_g^0$ (Figs. 3k and 3m). We note that the zero $\rho_{yx}$ plateau is absent in either uniformly doped or the Cr modulation doped QAH samples [23,26,31,32]. The 3-5-3 SH2 was magnetically trained first by an upward sweep up to $\mu_0H=1.5T$ before being swept downward. When $\mu_0H=-0.01T$, the MFM contrast is uniform (red), indicating that both top V- and bottom Cr-doped TI layers have upward magnetization (Fig. 3c). At $\mu_0H=-0.05T$, some reversed magnetic domains (green regions in Fig. 3d) appear, presumably in the 'softer' Cr-doped TI layer. As $\mu_0H$ is swept further, the green regions expand and fill up the whole scan area at $\mu_0H=-0.09T$, indicating the uniform antiparallel magnetization alignment over the entire 3-5-3 SH2. When $\mu_0H$ is further swept toward $H_{c1}$, new reversed magnetic domains (blue regions in Fig. 3g) nucleate at different locations, presumably in the 'harder' V-doped TI layer. Downward parallel magnetization alignment regions (blue) expand at the cost of antiparallel magnetization alignment regions (green) (Figs. 3g and 3h). When $\mu_0H=-1.0T$, the MFM image is uniformly blue, indicating a uniform downward parallel magnetization alignment (Fig. 3j). The direct observation of magnetic domains also allows us to extract the $H_c$s of magnetic layers. For our sandwich heterostructure, the $\mu_0H$ dependence of the magnetic domain contrast ($\delta f$), estimated from the root-mean-square (RMS) value of the MFM signal shows two peaks when the parallel and antiparallel magnetization alignment regions are equally populated (Fig. 3b). These two peaks correspond to the $H_c$s of top and bottom magnetic TI layers, in excellent agreement with those extracted from $\rho_{yx}$ shown in Fig. 3a. Our MFM measuments therefore demonstrated the presence of antiparallel magnetization alignment region (*i.e.* $-H_{c1} < H < -H_{c2}$) at 5.3K. It is reasonable to expect such behavior to persist to lower temperatures that bring out the QAH state.

Figure 3k shows the $\mu_0 H$ dependence of $\rho_{yx}$ of 3-5-3 SH2 over the temperature range 30mK$<T<$20K (see Fig. S9b for $\sigma_{xy}$). Both $\rho_{yx}$ and $\sigma_{xy}$ exhibit a two-step transition feature up to $T$=10K. The two-step feature sharpens with decreasing temperature showing zero $\rho_{yx}$ and $\sigma_{xy}$ plateaus for $T<$1.5K. To confirm the antiparallel magnetization alignment in zero $\rho_{yx}$ and $\sigma_{xy}$ plateau regions at $T$=30mK, we compared the minor loops at $T$=30mK with $V_g = V_g^0$ and $T$=5.3K with $V_g$=0V (Figs. 3l and 3m). The $H_E$ can be estimated by

$$H_E = \frac{\left| H_{c,\text{minor}}^L - H_{c,\text{minor}}^R \right|}{2},$$ where $H_{c,\text{minor}}^L$ and $H_{c,\text{minor}}^R$ denote the left and right $H_c$ of minor loops [34,35]. The $H_E$s at $T$=30mK and 5.3K are ~115Oe and ~84Oe, respectively. Since the $H_c$s of two magnetic TI layers are also functions of $T$, we define a relative $H_E$ value

$$\alpha = \frac{H_E}{H_{c1} - H_{c2}}$$ to compare the interlayer coupling strengths at different $T$s. We find that $\alpha$ ~0.013 at $T$=30mK and ~0.028 at $T$=5.3K. With the MFM observation at 5.3K, the smaller $\alpha$ at $T$=30mK confirms not surprisingly the robust antiparallel magnetization alignment in the zero $\rho_{yx}$ and $\sigma_{xy}$ plateau regions. Therefore, the axion insulator state is indeed realized in the zero $\rho_{yx}$ and $\sigma_{xy}$ plateau regions at $T$=30mK.

Flow diagram ($\sigma_{xy}$, $\sigma_{xx}$) of the 3-5-3 SH1 from $\mu_0 H$=1.8T to $\mu_0 H$=-1.8T (the red curves in Figs. 2a and 2b) at $T$=30mK is shown in Fig. 2c. Two circles of radius 0.5 $e^2/h$ centered at ($\sigma_{xy}$, $\sigma_{xx}$) = ($\pm 0.5 e^2/h$, 0) appear, the QAH state and the axion insulator state correspond to ($\sigma_{xy}$, $\sigma_{xx}$) = ($\pm e^2/h$, 0) and (0, 0), respectively. The sequential magnetization reversals of the top and bottom layers allow separation of the scaling of ($\sigma_{xy}$, $\sigma_{xx}$) of the top and bottom surfaces. In the antiparallel magnetization alignment (axion insulator) regions, the top V-doped TI layer maintains upward magnetization. When $\mu_0 H$ is tuned from 1.8T to $-0.45$T, the plateau transition around $-H_{c2} \sim -0.17$T is induced by the magnetization reversal of the Cr-doped TI layer, and the top surface keeps $(\sigma^t_{xy}, \sigma^t_{xx}) = (+0.5e^2/h, 0)$. Thus, we acquire the bottom surface scaling of $(\sigma^b_{xy}, \sigma^b_{xx}) = (\sigma_{xy}, \sigma_{xx}) - (\sigma^t_{xy}, \sigma^t_{xx})$, as shown in Fig. 2d, and a semicircle of radius $0.5e^2/h$ centered at (0,0) appears, consistent with the predicted flow diagram for a single surface state [3]. When $\mu_0 H$ is further tuned from -0.45T to -1.8T, the bottom surface contributes $(\sigma^b_{xy}, \sigma^b_{xx}) = (-0.5e^2/h, 0)$, so the top surface scaling of $(\sigma^t_{xy}, \sigma^t_{xx}) = (\sigma_{xy}, \sigma_{xx}) - (\sigma^b_{xy}, \sigma^b_{xx})$ is revealed (Fig. 2e). The same scaling behavior found at the top and bottom surfaces of the sandwich heterostructure confirms that each surface contributes half-integer quantization to the total $\sigma_{xy}$.

Figure 4 shows the $\mu_0 H$ dependence of $\rho_{xx}$, $\rho_{yx}$ and $\sigma_{xy}$ of 3-5-3 SH1 at various $V_g$s at 30mK. At $V_g = V_g^0 = +18$V, in the parallel magnetization alignment region, we find the QAH state with $\rho_{xx} \sim 126\Omega$ and $\rho_{yx}$ and $\sigma_{xy}$ fully quantized. However, in the antiparallel magnetization alignment region, $\rho_{xx}$ becomes very large (>1M$\Omega$) exceeding the range of our resistance bridge, and $\sigma_{xy}$ shows a zero plateau, corresponding to the axion insulator state. Unlike the $\rho_{yx}$ plateau in 3-5-3 SH2 (Figs. 3k and 3l), the $\rho_{yx}$ of 3-5-3 SH1 shows substantial fluctuations in the antiparallel magnetization alignment regions (see middle panels of Figs. 4b to 4d). This phenomenon is most likely due to the unavoidable electrical 'pickup' in $\rho_{yx}$ from the extremely large $\rho_{xx}$ with our imperfect Hall bar geometry.

The large $\rho_{xx}$ in antiparallel magnetization alignment confirms the insulating property of the axion insulator state. The top and bottom magnetic TI layers could be considered as a half-integer 'QAH insulator' with $\sigma_{xy}=0.5e^2/h$, but the 1D chiral edge mode of the entire sandwich heterostructure is forbidden due to their antiparallel magnetization alignment. Furthermore, the observed large $\rho_{xx}$ also exclude the existence of helical side surface states in the sandwich heterostructure, reflecting the effect of quantum confinement. Since there is no conduction channel in the 3-5-3 sandwich heterostructure with antiparallel magnetization alignment, the axion insulator state is much more insulating than that induced by scattering from multi-magnetic domains in uniformly magnetically doped QAH samples [10-15,25,31,32]. Thus, the external $\mu_0 H$ can turn OFF/ON the 1D chiral edge state in antiparallel/parallel magnetization alignment regions, leading to an 'OFF/ON' ratio of ~8000 for $\rho_{xx}$.

In contrast to Ref. [23], we observe zero $\rho_{yx}$ plateau in addition to zero $\sigma_{xy}$ plateau in our electrical transport measurements. The zero $\sigma_{xy}$ and $\rho_{yx}$ plateaus are found in the antiparallel magnetization alignment region, as confirmed by magnetic force microscopy (MFM) measurements. These observations exclude the possibility that zero $\sigma_{xy}$ plateau in our QAH sandwich heterostructure is an arithmetical artifact due to a large $\rho_{xx}$ at $H_c$.

The QAH and axion insulator states persist when *n*-type carriers are introduced (Figs. 4c to 4e). The $\rho_{xx}$ in antiparallel magnetization alignment is tuned from $>40h/e^2$ at $V_g=V_g^0$ to $~3h/e^2$ at $V_g=+200$V, and the zero $\rho_{yx}$ plateau becomes more obvious, as shown in the middle panel of Fig. 4e. When $V_g<0$V, $\rho_{xx}$, $\rho_{yx}$ and $\sigma_{xy}$ are significantly altered, and their values become very small (Fig. 4a), indicating that the chemical potential of the sandwich heterostructure has crossed the bulk valence bands. In this case, the FM order is modulated by the Ruderman-Kittel-Kasuya-Yosida interaction [36], rather than the van Vleck interaction when the chemical potential is located in the vicinity of $V_g=V_g^0$ [8].

To summarize, our transport studies, complemented by MFM measurements, demonstrated the axion insulator state in the antiparallel magnetization alignment of the top and bottom magnetic layers of the QAH sandwich heterostructures. By separately analyzing the scaling of the transitions of the top and bottom magnetic layers, we show clearly that each surface of the QAH insulator or axion insulator contributes half-integer quantization. The demonstration of the axion insulator state in QAH sandwich heterostructures paves the way for the experimental exploration of the TME effect and other fascinating topological phenomena.

*Note added:* During the review process of this paper, we became aware of similar work published in *Sci. Adv.* [37].

**Acknowledgements**

The authors would like to thank K. F. Mak, A. H. MacDonald and B. H. Yan for the helpful discussions and A. Richardella, T. Pillsbury, J. Kally, W. Zhao, Z. Chen, H. L. Fu, X. Lin, S. Jiang and S. Kempinger for help with experiments. D. X., J. J., F. W., Y. Z., C. L., C. Z. C. and N. S. acknowledge support from the Penn State Two-Dimensional Crystal Consortium-Materials Innovation Platform (2DCC-MIP) under NSF grant DMR-1539916. D. X. and N. S. also acknowledge support from Office of Naval Research (Grant No. N00014-15-1-2370) and from ARO MURI (W911NF-12-1-0461). J. H. S. and M. H. W. C. acknowledge the support from NSF grant DMR-1707340. C. X. L acknowledges the support from Office of Naval Research (Grant No. N00014-15-1-2675). C. Z. C. acknowledges support from a startup grant provided by Penn State. Work at Rutgers was supported by the U.S. Department of Energy (DOE), Office of Science, Basic Energy Sciences (BES) under Award # DE-SC0018153.

## References

[1] X. L. Qi and S. C. Zhang, *Rev. Mod. Phys.* **83**, 1057 (2011).

[2] M. Z. Hasan and C. L. Kane, *Rev. Mod. Phys.* **82**, 3045 (2010).

[3] X. L. Qi, T. L. Hughes, and S. C. Zhang, *Phys. Rev. B* **78**, 195424 (2008).

[4] J. Wang, B. Lian, X. L. Qi, and S. C. Zhang, *Phys. Rev. B* **92**, 081107 (2015).

[5] T. Morimoto, A. Furusaki, and N. Nagaosa, *Phys. Rev. B* **92**, 085113 (2015).

[6] A. M. Essin, J. E. Moore, and D. Vanderbilt, *Phys. Rev. Lett.* **102**, 146805 (2009).

[7] F. Wilczek, *Phys. Rev. Lett.* **58**, 1799 (1987).

[8] R. Yu, W. Zhang, H. J. Zhang, S. C. Zhang, X. Dai, and Z. Fang, *Science* **329**, 61 (2010).

[9] C. X. Liu, X. L. Qi, X. Dai, Z. Fang, and S. C. Zhang, *Phys. Rev. Lett.* **101**, 146802 (2008).

[10] C. Z. Chang, J. S. Zhang, X. Feng, J. Shen, Z. C. Zhang, M. H. Guo, K. Li, Y. B. Ou, P. Wei, L. L. Wang, Z. Q. Ji, Y. Feng, S. H. Ji, X. Chen, J. F. Jia, X. Dai, Z. Fang, S. C. Zhang, K. He, Y. Y. Wang, L. Lu, X. C. Ma, and Q. K. Xue, *Science* **340**, 167 (2013).

[11] X. F. Kou, S. T. Guo, Y. B. Fan, L. Pan, M. R. Lang, Y. Jiang, Q. M. Shao, T. X. Nie, K. Murata, J. S. Tang, Y. Wang, L. He, T. K. Lee, W. L. Lee, and K. L. Wang, *Phys. Rev. Lett.* **113**, 137201 (2014).

[12] J. G. Checkelsky, R. Yoshimi, A. Tsukazaki, K. S. Takahashi, Y. Kozuka, J. Falson, M. Kawasaki, and Y. Tokura, *Nat. Phys.* **10**, 731 (2014).

[13] C. Z. Chang, W. W. Zhao, D. Y. Kim, H. J. Zhang, B. A. Assaf, D. Heiman, S. C. Zhang, C. X. Liu, M. H. W. Chan, and J. S. Moodera, *Nat. Mater.* **14**, 473 (2015).

[14] M. H. Liu, W. D. Wang, A. R. Richardella, A. Kandala, J. Li, A. Yazdani, N. Samarth, and N. P. Ong, *Sci. Adv.* **2**, e1600167 (2016).

[15] A. Kandala, A. Richardella, S. Kempinger, C. X. Liu, and N. Samarth, *Nat. Commun.* **6**, 7434 (2015).

[16] W. K. Tse and A. H. MacDonald, *Phys. Rev. Lett.* **105**, 057401 (2010).

[17] J. Maciejko, X. L. Qi, H. D. Drew, and S. C. Zhang, *Phys. Rev. Lett.* **105**, 166803 (2010).

[18] K. Nomura and N. Nagaosa, *Phys. Rev. Lett.* **106**, 166802 (2011).

[19] X. L. Qi, R. D. Li, J. D. Zang, and S. C. Zhang, *Science* **323**, 1184 (2009).

[20] L. Wu, M. Salehi, N. Koirala, J. Moon, S. Oh, and N. P. Armitage, *Science* **354**, 1124 (2016).

[21] V. Dziom, A. Shuvaev, A. Pimenov, G. V. Astakhov, C. Ames, K. Bendias, J. Bottcher, G. Tkachov, E. M. Hankiewicz, C. Brune, H. Buhmann, and L. W. Molenkamp, *Nat. Commun.* **8**, 15197 (2017).

[22] K. N. Okada, Y. Takahashi, M. Mogi, R. Yoshimi, A. Tsukazaki, K. S. Takahashi, N. Ogawa, M. Kawasaki, and Y. Tokura, *Nat. Commun.* **7**, 12245 (2016).

[23] M. Mogi, M. Kawamura, R. Yoshimi, A. Tsukazaki, Y. Kozuka, N. Shirakawa, K. S. Takahashi, M. Kawasaki, and Y. Tokura, *Nat. Mater.* **16**, 516 (2017).

[24] E. O. Lachman, M. Mogi, J. Sarkar, A. Uri, K. Bagani, Y. Anahory, Y. Maysoedov, M. E. Huber, A. Tsukazaki, M. Kawasaki, Y. Tokura, and E. Zeldov, *arXiv*:1710.06446 (2017).

[25] S. Grauer, K. M. Fijalkowski, S. Schreyeck, M. Winnerlein, K. Brunner, R. Thomale, C. Gould, and L. W. Molenkamp, *Phys. Rev. Lett.* **118**, 246801 (2017).

[26] M. Mogi, R. Yoshimi, A. Tsukazaki, K. Yasuda, Y. Kozuka, K. S. Takahashi, M. Kawasaki, and Y. Tokura, *Appl. Phys. Lett.* **107**, 182401 (2015).

[27] K. Yasuda, R. Wakatsuki, T. Morimoto, R. Yoshimi, A. Tsukazaki, K. S. Takahashi, M. Ezawa, M. Kawasaki, N. Nagaosa, and Y. Tokura, *Nat. Phys.* **12**, 555 (2016).

[28] Y. Zhang, K. He, C. Z. Chang, C. L. Song, L. L. Wang, X. Chen, J. F. Jia, Z. Fang, X. Dai, W. Y. Shan, S. Q. Shen, Q. Niu, X. L. Qi, S. C. Zhang, X. C. Ma, and Q. K. Xue, *Nat. Phys.* **6**, 584 (2010).

[29] Y. P. Jiang, Y. L. Wang, M. Chen, Z. Li, C. L. Song, K. He, L. L. Wang, X. Chen, X. C. Ma, and Q. K. Xue, *Phys. Rev. Lett.* **108**, 016401 (2012).

[30] T. Zhang, J. Ha, N. Levy, Y. Kuk, and J. Stroscio, *Phys. Rev. Lett.* **111**, 056803 (2013).

[31] X. F. Kou, L. Pan, J. Wang, Y. B. Fan, E. S. Choi, W. L. Lee, T. X. Nie, K. Murata, Q. M. Shao, S. C. Zhang, and K. L. Wang, *Nat. Commun.* **6**, 8474 (2015).

[32] Y. Feng, X. Feng, Y. B. Ou, J. Wang, C. Liu, L. G. Zhang, D. Y. Zhao, G. Y. Jiang, S. C. Zhang, K. He, X. C. Ma, Q. K. Xue, and Y. Y. Wang, *Phys. Rev. Lett.* **115**, 126801 (2015).

[33] J. Wang, B. Lian, and S. C. Zhang, *Phys. Rev. B* **89**, 085106 (2014).

[34] P. Walser, M. Hunziker, T. Speck, and M. Landolt, *Phys. Rev. B* **60**, 4082 (1999).

[35] J. Faure-Vincent, C. Tiusan, C. Bellouard, E. Popova, M. Hehn, F. Montaigne, and A. Schuhl, *Phys. Rev. Lett.* **89**, 107206 (2002).

[36] Z. C. Zhang, X. Feng, M. H. Guo, K. Li, J. S. Zhang, Y. B. Ou, Y. Feng, L. L. Wang, X. Chen, K. He, X. C. Ma, Q. K. Xue, and Y. Y. Wang, *Nat. Commun.* **5**, 4915 (2014).

[37] M. Mogi, M. Kawamura, A. Tsukazaki, R. Yoshimi, K. S. Takahashi, M. Kawasaki, and Y. Tokura, *Sci. Adv.* **3**, eaao1669 (2017).

**Figures and figure captions:**

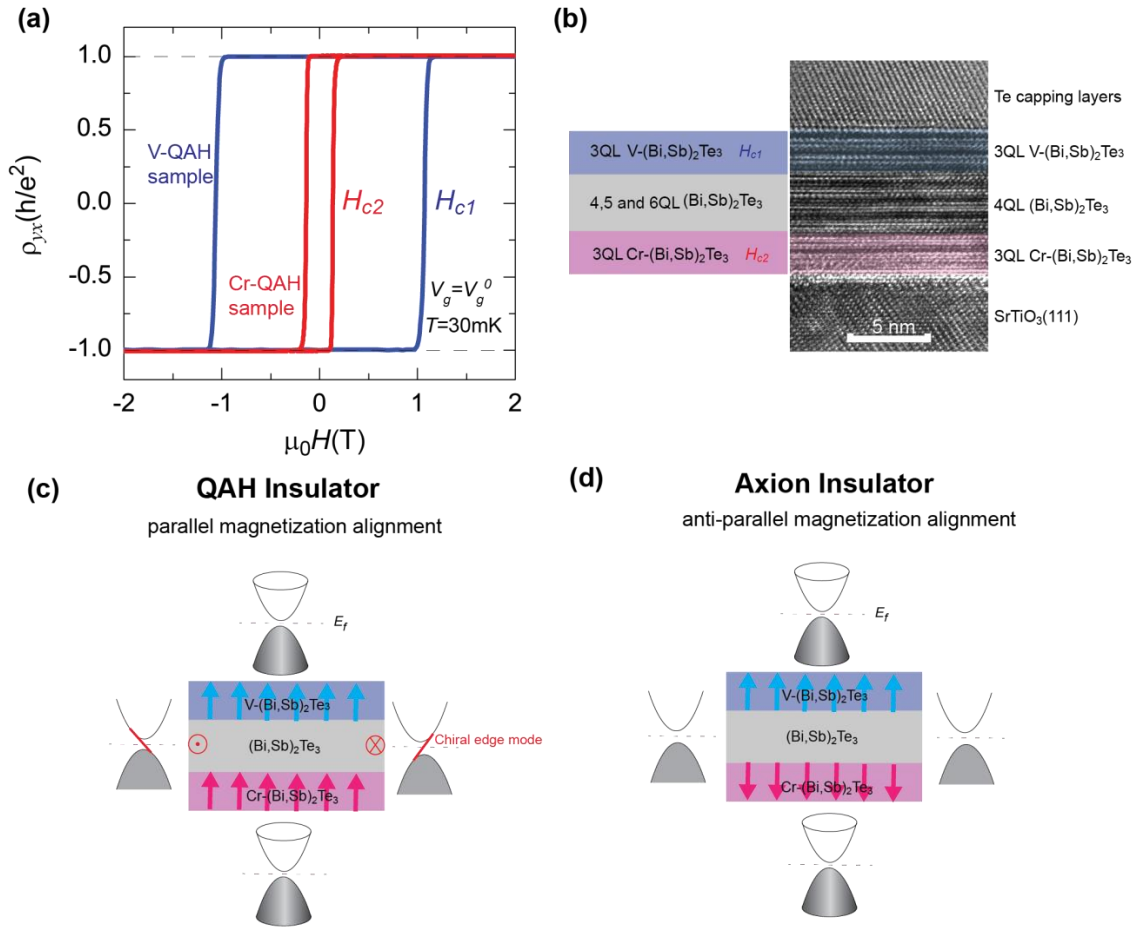

**Figure 1| V-doped (Bi,Sb)$_2$Te$_3$/TI/Cr-doped (Bi,Sb)$_2$Te$_3$ sandwich heterostructure.** (a) Signatures of the QAH insulator states in individual Cr- and V-doped (Bi,Sb)$_2$Te$_3$ films measured at $T$=30mK. The $H_c$ of the V-doped QAH film $H_{c1}$~1.06T; the $H_c$ of the Cr-doped QAH film $H_{c2}$~0.14T. (b) Schematic drawing and cross-sectional TEM image of a QAH sandwich heterostructure. (c) Schematics of a sandwich heterostructure with parallel magnetization alignment surfaces, supporting a 1D chiral edge state. (d) Schematics of a sandwich heterostructure with antiparallel magnetization alignment surfaces, supporting an axion insulator state.

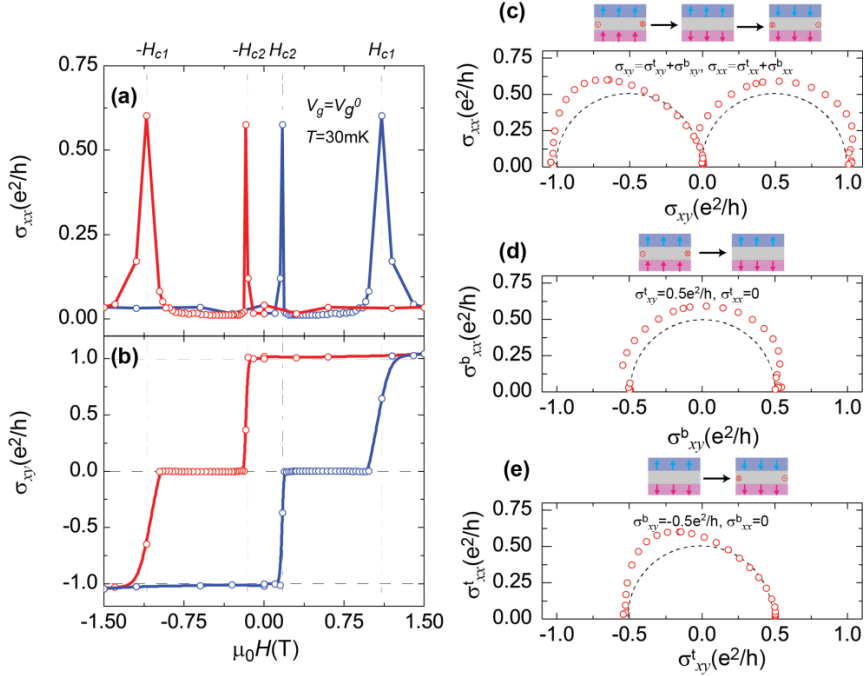

**Figure 2| Signature of an axion insulator in a 3-5-3 SH1 with antiparallel magnetization alignment.** (a, b) $\mu_0 H$ dependence of $\sigma_{xx}$ and $\sigma_{xy}$ of 3-5-3 SH1 at $V_g=V_g^0$ and $T$=30mK. An axion insulator is observed if the zero $\sigma_{xy}$ plateau region is the consequence of antiparallel magnetization alignment. The red and blue lines are guides to the eye. (c) Flow diagram of ($\sigma_{xy}$, $\sigma_{xx}$) of the 3-5-3 SH1 from $\mu_0H$=1.8T to $\mu_0H$=-1.8T (the red curves in (a) and (b)). Two semicircles of radius $0.5h/e^2$ centered at ($0.5h/e^2$, 0) and ($-0.5h/e^2$, 0) are shown in dashed lines. (d) Flow diagram of ($\sigma^b_{xy}$, $\sigma^b_{xx}$) of the bottom surface from $\mu_0H$=1.8T to $\mu_0H$=-0.45T. (e) Flow diagram of ($\sigma^t_{xy}$, $\sigma^t_{xx}$) of the top surface from $\mu_0H$=-0.45T to $\mu_0H$=-1.8T. A semicircle of radius $0.5h/e^2$ centered at (0, 0) is shown in a dashed line in (d) and (e). Here $\mu_0H$=-0.45T is chosen to separately study the scalings of the top and bottom surfaces of the QAH sandwich heterostructures in proper magnetic field sweeping regions.

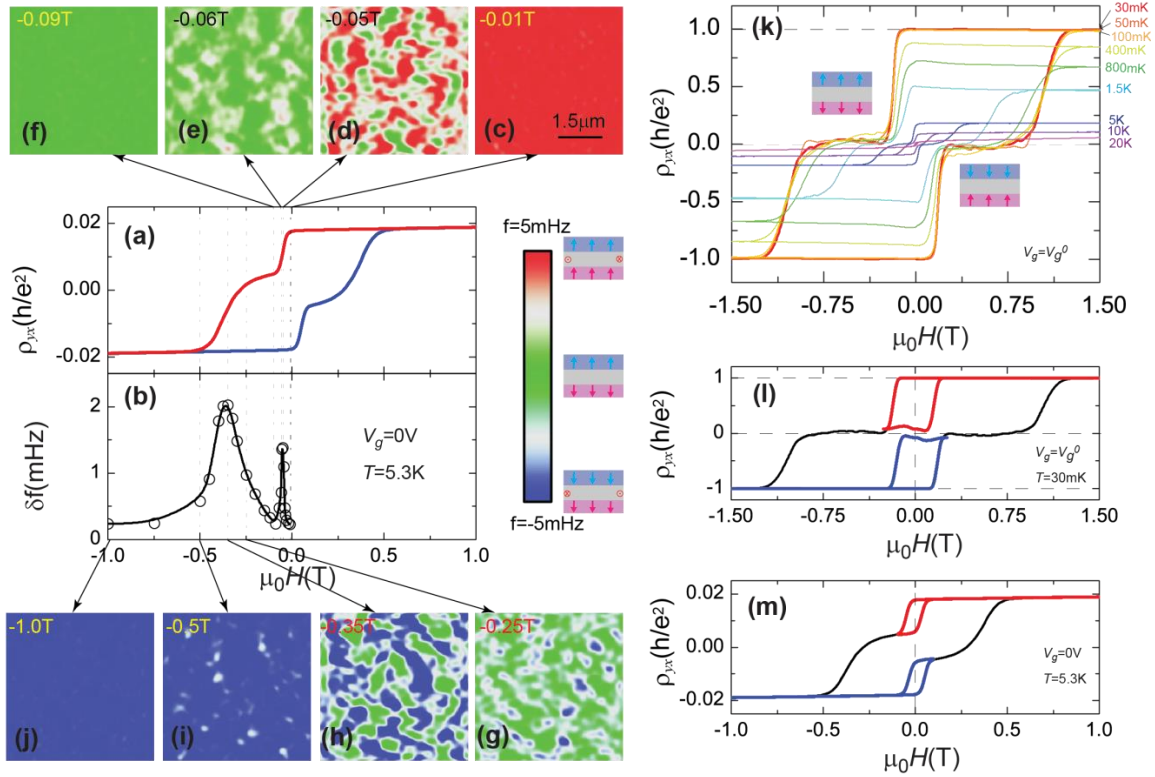

**Figure 3| Two-step magnetization reversal in the 3-5-3 SH2.** (a) $\mu_0H$ dependence of $\rho_{xy}$ of 3-5-3 SH2 at $V_g=0V$ and $T=5.3K$. (b) $\mu_0H$ dependence of magnetic domain contrasts ($\delta f$) at $V_g=0$ and $T=5.3K$. The $\delta f$ is estimated from the root-mean-square (RMS) value of MFM images in (c-j). (c-j) The MFM images at various $\mu_0H$s. Red and blue represent respectively upward and downward parallel magnetization alignment regions, green represents antiparallel magnetization alignment regions. The Hall traces in (a) and the MFM images in (c-j) are measured simultaneously. (k) $\mu_0H$ dependence of $\rho_{yx}$ at various temperatures. $\rho_{yx}$ exhibits two-step transition when $T<10K$. (l, m) Minor loops of the $\rho_{yx}$ at $V_g=V_g^0$ and $T=30mK$ (l) and at $V_g=0V$ and $T=5.3K$ (m).

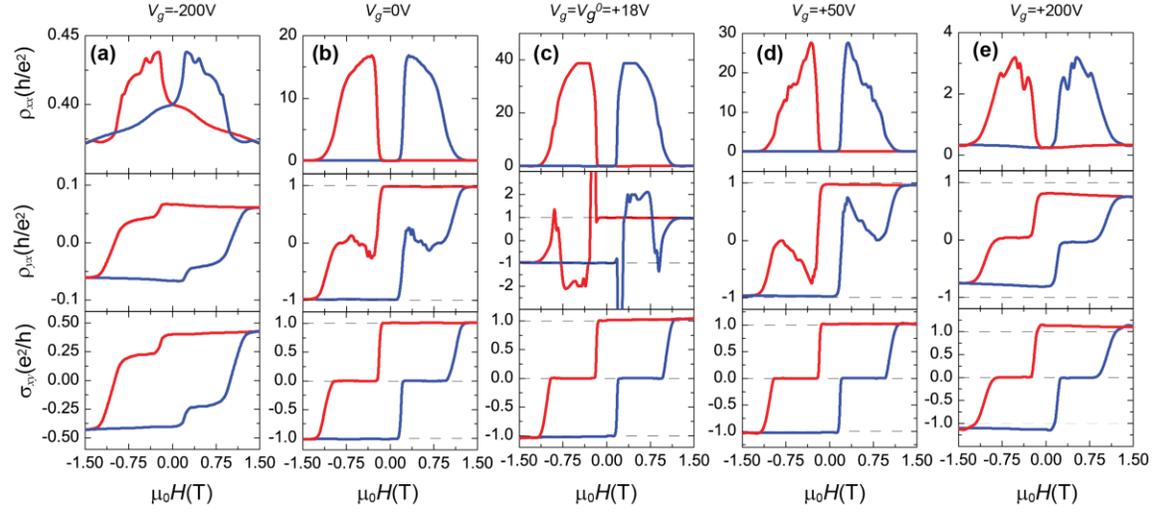

**Figure 4| Gate-tuning of $\rho_{xx}$, $\rho_{yx}$ and $\sigma_{xy}$ of the 3-5-3 SH1 at $T$=30mK.** (a-e) $\mu_0H$ dependence of $\rho_{xx}$ (top panels), $\rho_{yx}$ (middle panels) and $\sigma_{xy}$ (bottom panels) at varied $V_g$s. The red and blue curves show results sweeping $\mu_0H$ downward and upward, respectively.

Supporting Materials for

# Realization of the Axion Insulator State in Quantum Anomalous Hall Sandwich Heterostructures

Di Xiao[1]*, Jue Jiang[1]*, Jae-Ho Shin[1], Wenbo Wang[2], Fei Wang[1], Yi-Fan Zhao[1], Chaoxing Liu[1], Weida Wu[2], Moses H. W. Chan[1], Nitin Samarth[1] and Cui-Zu Chang[1]

[1]Department of Physics, Pennsylvania State University, University Park, PA16802

[2]Department of Physics and Astronomy, Rutgers University, Piscataway, NJ 08854

*These authors contributed equally to this work.

Corresponding authors: mhc2@psu.edu (M.H.W.C.); nxs16@psu.edu (N.S.); cxc955@psu.edu (C.Z.C.)

**Contents:**

## I. MBE growth and characterizations of sandwich heterostructures

The insulating SrTiO$_3$(111) substrates used for the growth of all the sandwich heterostructures were first soaked in 90°C deionized water for 1.5 hours, and then annealed at 985 °C for 3 hours in a tube furnace with flowing pure oxygen gas. Through the above heat treatments, the SrTiO$_3$(111) substrate surface become passivated and atomically flat, the topological insulator (TI) sandwich heterostructure growth was carried out using a commercial EPI-620 molecular beam epitaxy (MBE) system with a vacuum that is better than $2\times10^{-10}$ mbar. The heat-treated insulating SrTiO$_3$ (111) substrates were outgassed at ~530 °C for 1 hour before the growth of the TI sandwich heterostructures. High-purity Bi (99.999%), Sb (99.9999%), Cr (99.999%) and Te (99.9999%) were evaporated from Knudsen effusion cells, and V (99.995%) was evaporated from an e-gun. During growth of the TI, the substrate was maintained at ~240 °C. The flux ratio of Te per (Bi + Sb) was set to be >10 to prevent Te deficiency in the samples. The pure or magnetic TI growth rate was at ~0.25 QL/min. Each layer of the sandwich heterostructure was grown with the different Bi/Sb ratio by adjusting their K-cell temperatures to tune the chemical potential close to its charge neutral point. Following the growth, the TI films were annealed at ~240 °C for 30 minutes to improve the crystal quality before being cooled down to room temperature. Finally, to avoid possible contamination, an 18 nm thick Te layer is deposited at room temperature on top of the sandwich heterostructures prior to their removal from the MBE chamber for transport measurements.

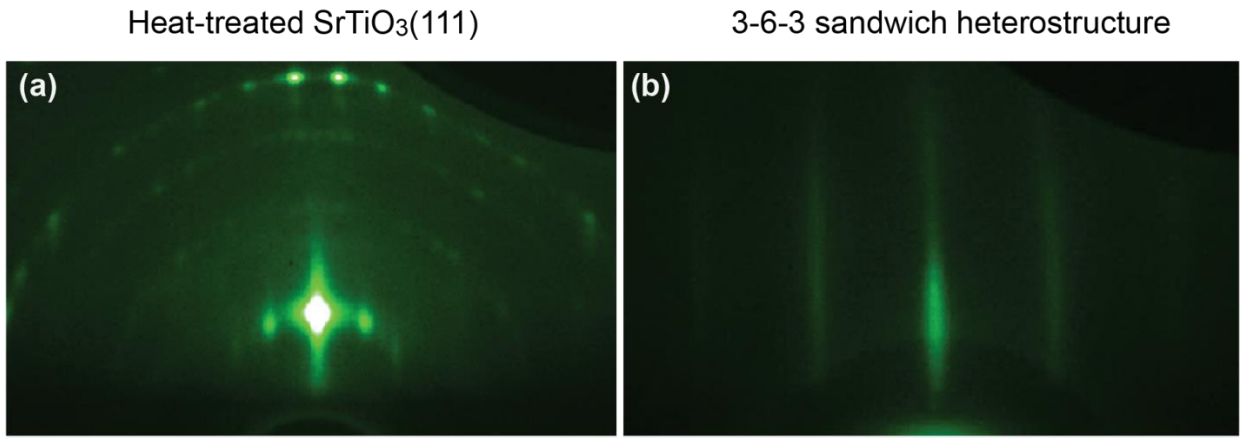

**Figure S1| RHEED patterns of the substrate and the sandwich heterostructure.** (a,b) RHEED patterns of the heat-treated SrTiO$_3$ (111) substrate (a) and a 3-6-3 sandwich heterostructure (b).

Figure S1 displays the reflective high energy electron diffraction (RHEED) patterns of a heat-treated bare insulating SrTiO$_3$ (111) substrate and a 3-6-3 quantum anomalous Hall (QAH) sandwich heterostructure. The clear reconstruction of the heat-treated SrTiO$_3$ (111) substrate (Fig. S1a) indicates its atomic flat surface, which is suitable for the high-quality QAH sandwich heterostructure growth. The sharp and streaky '1×1' patterns of the 3-6-3 QAH sandwich heterostructure (Fig. S1b) are signatures of the highly-ordered and smooth surface.

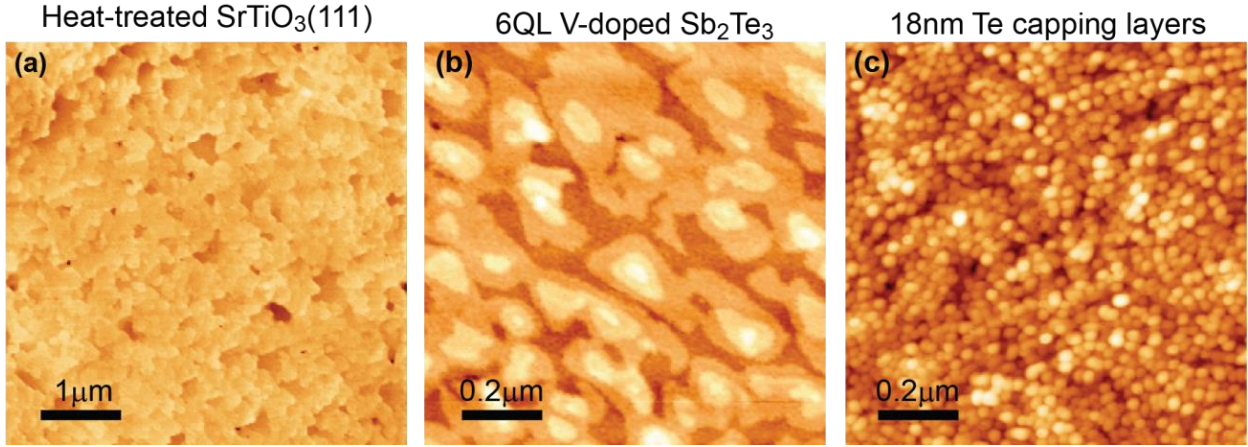

**Figure S2|  AFM images of the substrate, the magentically doped TI films and the capping layers.** (a) The AFM image of a heat-treated $SrTiO_3$ (111) surface with root-mean-square (RMS) roughness $R_q$ = 0.33 nm. (b) The AFM image of a 6QL V-doped $Sb_2Te_3$ thin film grown on $SrTiO_3$ (111) surface with $R_q$ = 0.70 nm. (c) The AFM image of 18 nm thick Te layers on the 3-6-3 sandwich heterostruture with $R_q$ = 1.0 nm.

Figure S2 shows the atomic force microscope (AFM) images of a typical heat-treated $SrTiO_3$ (111) substrate (Fig. S2a), a 6 QL V-doped $Sb_2Te_3$ grown on the heat-treated $SrTiO_3$ (111) substrate (Fig. S2b) and the 18 nm Te layers on the 3-6-3 sandwich heterostruture (Fig. S2c). The root-mean-square (RMS) roughness ($R_q$) of both the $SrTiO_3$ (111) substrate and the 6QL V-doped $Sb_2Te_3$ film is < 1nm, indicating the atomically flat surfaces of the substrate and the magnetically doped TI films. The 18 nm thick Te capping layer is a good insulator with the sheet longitudinal resistance $R_{xx}$ >100 MΩ at cryogenic temperature [1], which is several orders of magnitude larger than that of the sandwich heterostructure. Therefore, the 18 nm Te capping layer has no effect on the transport properties of the sandwich heterostructures.

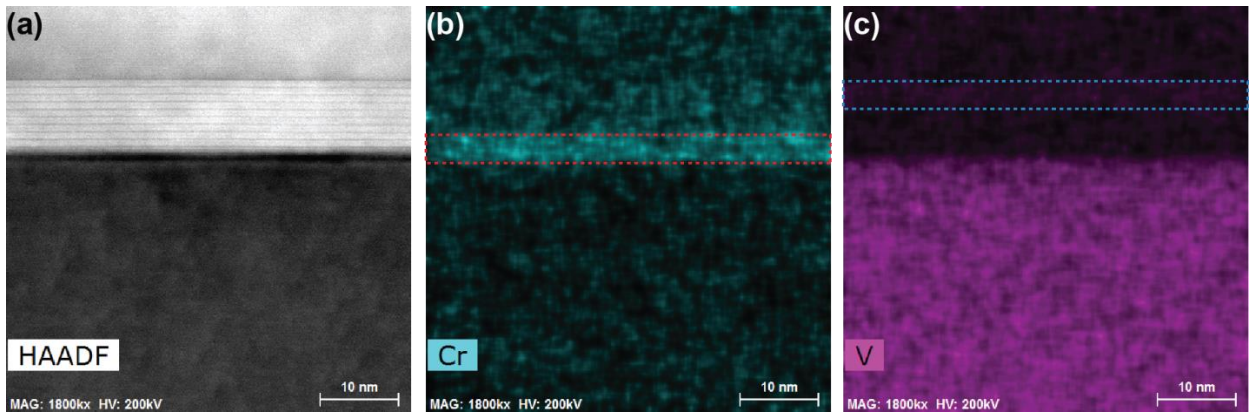

**Figure S3| TEM image and EDS mappings of the 3-4-3 sandwich heterostructure.** (a) The cross-sectional TEM image of 3-4-3 heterostructure with a total thickness of ~10QL. (b, c) The EDS mappings of Cr (b) and V (c). The Cr (cyan) and V signals (purple) concentrate in the bottom 3QL and top 3QL of the sandwich heterostructure. The strong purple color in the $SrTiO_3$ (111) substrate is from the Ti atoms, whose peak is close to that of the V atoms.

Figure S3 shows the cross-sectional transmission electron microscopy (TEM) image of the 3-4-3 sandwich heterostructure and the corresponding energy-dispersive X-ray spectroscopy (EDS) mappings. The cross-sectional TEM image in Fig. 3a confirms the good crystal quality of the sandwich heterostructure. The TEM results also show the total thickness of 3-4-3 sandwich heterostructure is ~10 quintuple layers (QL). The EDS mappings of the Cr and V atom signals in the cross section (Figs. S3b and S3c) show the bottom and top layers of the 3-4-3 sandwich heterostructure are separately doped with Cr and V. The weak Cr and V signals in bottom and top layers reflect the diluted doping (~5%) levels of these ions in our sandwich heterostructure. The strong purple signal observed in the $SrTiO_3$ (111) substrate, as shown in Fig. S3c, is from the Ti atoms rather than V atoms. This is because of the similar peak positions of V and Ti atoms in their EDS spectra.

## II. Device fabrications and results from PPMS cryostat

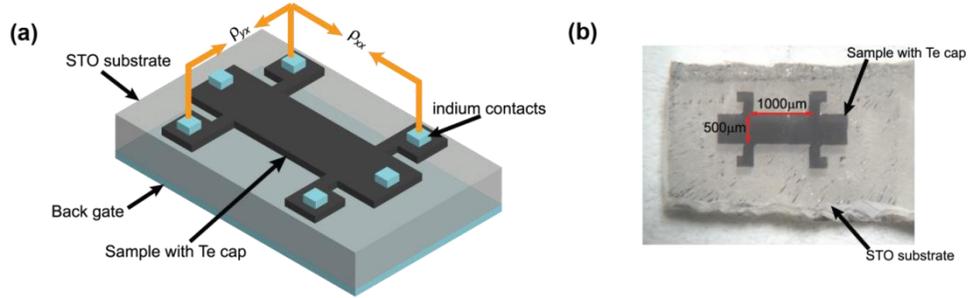

**Figure S4| Hall bar devices in transport measurements.** (a) Schematics of the Hall bar devices. (b) The photograph of a Hall bar device used in the transport measurements.

All the sandwich heterostructures in our transport measurements are scratched into Hall bar geometry (Fig. S4a) using a computer controlled probe station. The effective area of the Hall bar device is ~ 1 × 0.5 mm. The electrical contacts for transport measurements are made by pressing tiny indium spheres on the Hall bar. The bottom gate electrode is prepared through an indium foil on the back side of the $SrTiO_3$ substrate. The real Hall bar device is shown in Fig. S4b. More than 20 sandwich heterostructure devices are screened by a Physical Property Measurement System (PPMS) cryostat (Quantum Design, 2 K, 9 T). A dilution refrigerator (Leiden Cryogenics, 10 mK, 9 T) is used to carry lower temperature (30 mK < $T$ < 2 K) transport measurements on a total of six sandwich heterostructures with the near ideal chemical compositions.

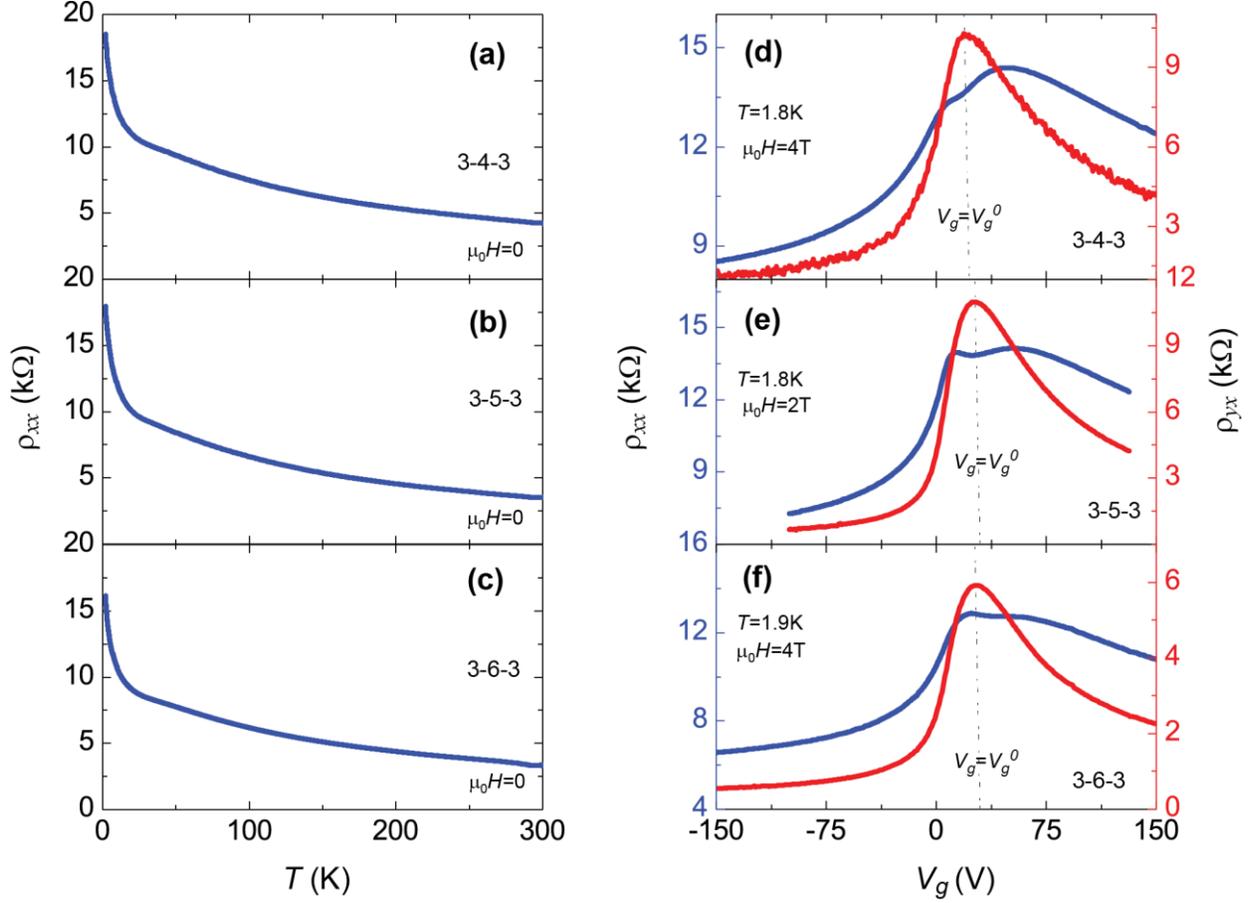

**Figure S5| PPMS results of the three sandwich heterostructure sample wafers that were subsequently studied in the Leiden Fridge.** (a-c) The $T$ dependence of $\rho_{xx}$ for 3-4-3 (a), 3-5-3 (b), and 3-6-3 (c) sandwich heterostructures. (d-f) The $V_g$ dependence of $\rho_{xx}$ (blue) and $\rho_{yx}$ (red) for 3-4-3 (d), 3-5-3 (e), and 3-6-3 (f) sandwich heterostructures at $T$=2K.

Figure 5 shows the PPMS results of the three sandwich heterostructures with the undoped TI spacer thickness of 4 QL, 5 QL and 6 QL. The $T$ dependence of $\rho_{xx}$ for 3-4-3, 3-5-3 and 3-6-3 sandwich heterostructures is shown in Figs. S5a to S5c. Without magnetic training, the $\rho_{xx}$ increases with decreasing $T$ for all the three sandwich heterostructures, suggesting their insulating behaviors and indicating that the chemical potential in each layer of the sandwich heterostructure is close to its charge neutral point ($V_g^0$). Figs. S5d to S5f show the gate ($V_g$) dependence of the $\rho_{xx}$ and $\rho_{yx}$ for 3-4-3, 3-5-3, and 3-6-3 sandwich heterostructures at $T = 2$ K, respectively. In all the three sandwich heterostructures, when the $\rho_{yx}$ shows a peak, the $\rho_{xx}$ shows a dip at $V_g=V_g^0$. This phenomenon signals the appearance of the QAH state at $T = 2$ K [2,3]. These three sandwich heterostructure sample wafers were selected for systematic lower temperature transport measurements in the Leiden fridge.

## III. Gate dependence of magneto-transport results of the three sandwich heterostructures

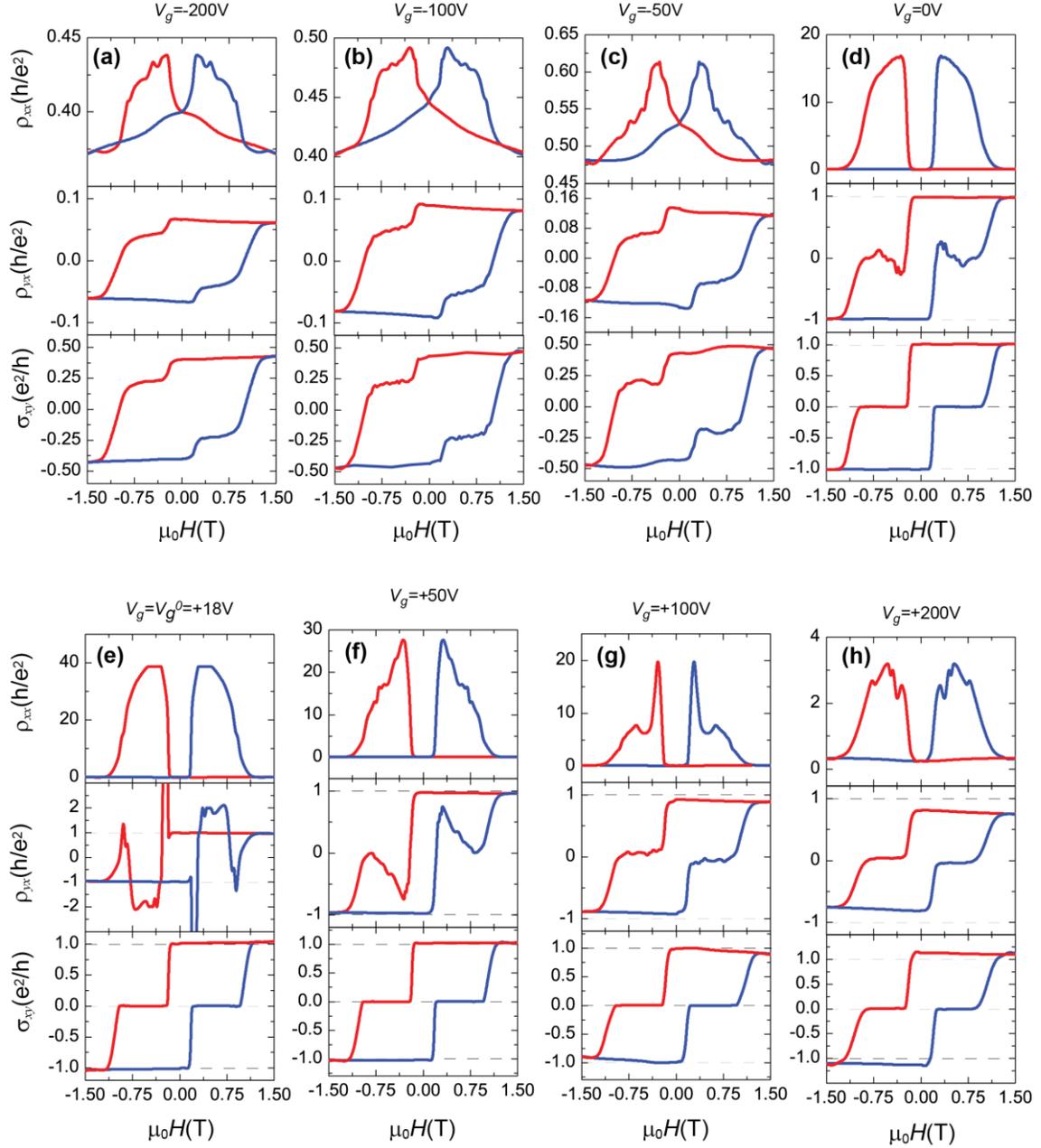

**Figure S6| Gate-tuning of $\rho_{xx}$, $\rho_{yx}$ and $\sigma_{xy}$ of the 3-5-3 SH1 at $T$ = 30 mK.** (a-h) $\mu_0H$ dependence of $\rho_{xx}$ (top panels), $\rho_{yx}$ (middle panels) and $\sigma_{xy}$ (bottom panels) at varied $V_g$s. The red and blue curves show results sweeping $\mu_0H$ downward and upward, respectively.

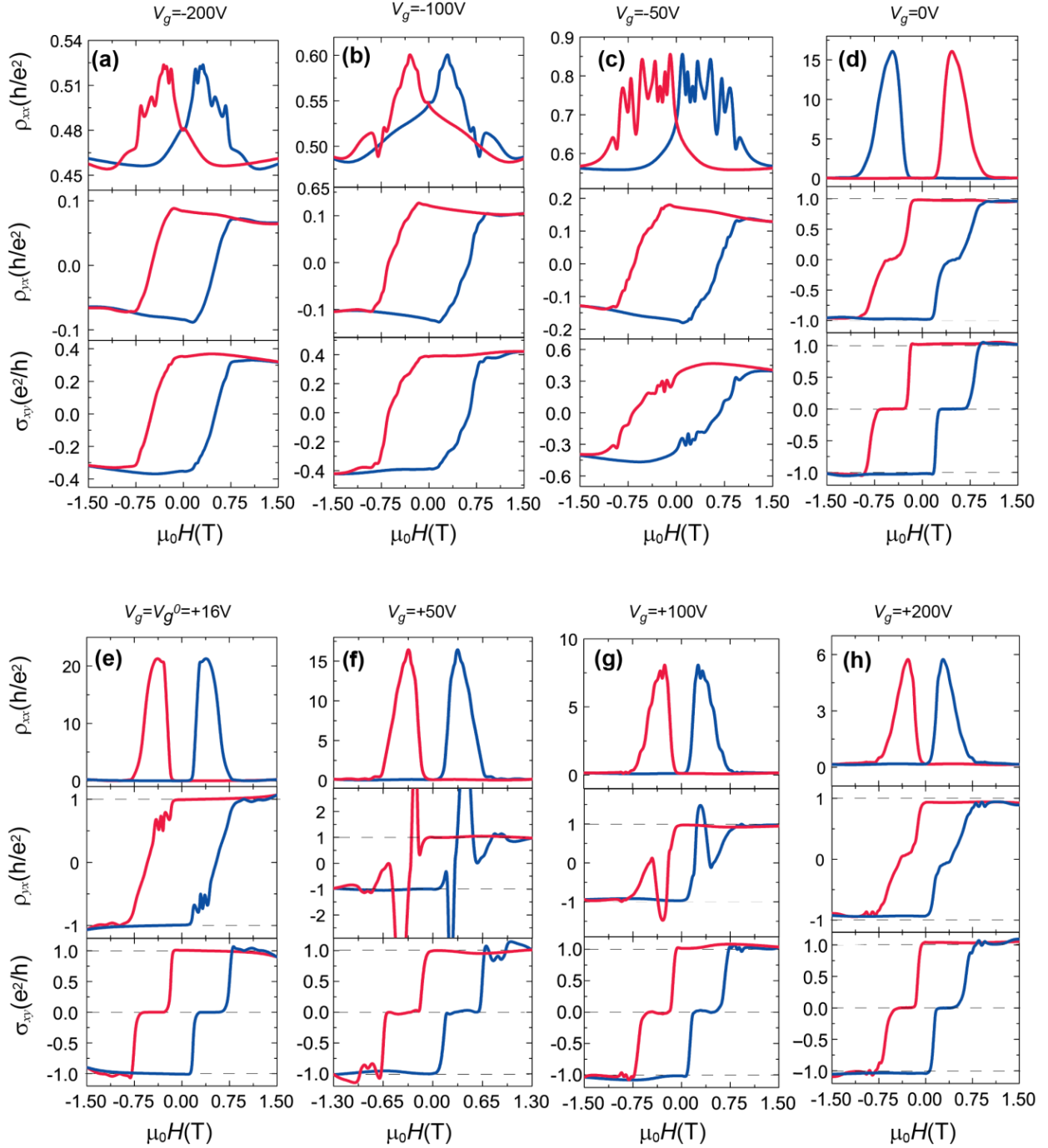

**Figure S7| Gate-tuning of $\rho_{xx}$, $\rho_{yx}$ and $\sigma_{xy}$ of the 3-4-3 sandwich heterostructure at $T = 30$ mK.** (a-h) $\mu_0H$ dependence of $\rho_{xx}$ (top panels), $\rho_{yx}$ (middle panels) and $\sigma_{xy}$ (bottom panels) at varied $V_g$s. The red and blue curves show results sweeping $\mu_0H$ downward and upward, respectively.

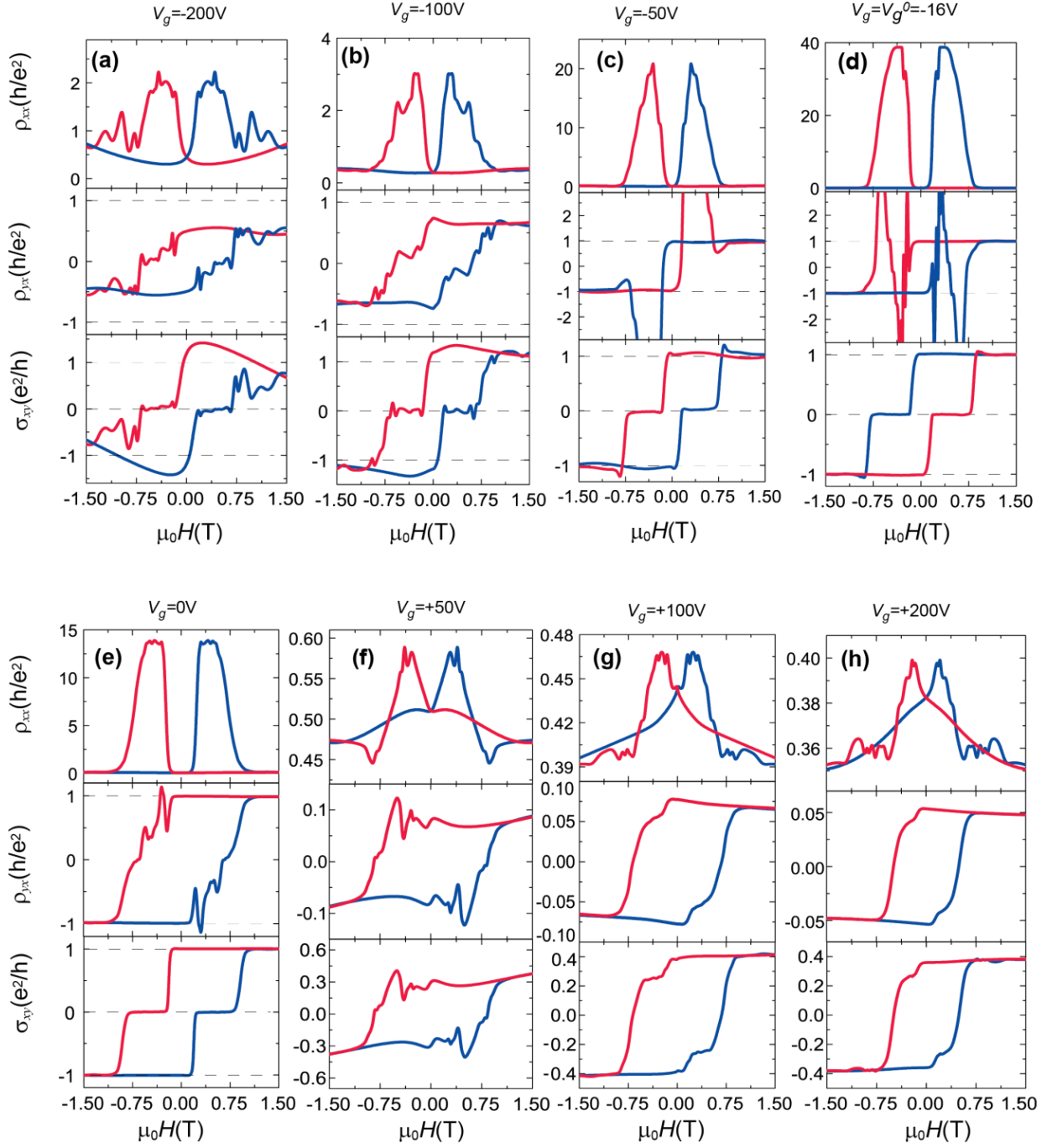

**Figure S8| Gate-tuning of $\rho_{xx}$, $\rho_{yx}$ and $\sigma_{xy}$ of the 3-6-3 sandwich heterostructure at $T$ = 30 mK.** (a-h) $\mu_0H$ dependence of $\rho_{xx}$ (top panels), $\rho_{yx}$ (middle panels) and $\sigma_{xy}$ (bottom panels) at varied $V_g$s. The red and blue curves show results sweeping $\mu_0H$ downward and upward, respectively.

We systematically carried the $V_g$ dependence measurements on the three sandwich heterostructures with different spacer thicknesses (3-4-3, 3-5-3 (SH1) and 3-6-3 sandwich heterostructures) at $T$=30 mK. A small excitation current of ~0.3nA flowing through the film plane was used in the transport measurements to minimize any heating effect, and a perpendicular $\mu_0H$ was used to adjust the magnetizations of top and bottom magnetic layers in our transport measurements.

Figures S6 to S8 show the $\mu_0H$ dependence of $\rho_{xx}$, $\rho_{yx}$ and $\sigma_{xy}$ at various $V_g$s for 3-5-3(SH1) (Fig. S6), 3-4-3 (Fig. S7), and 3-6-3 (Fig. S8) sandwich heterostructures. The $V_g^0$'s of 3-5-3(SH1), 3-4-3 and 3-6-3 sandwich heterostructures are 18 V, 16 V, and -16 V, respectively. The three sandwich heterostructures show very similar results. At $V_g = V_g^0$, the QAH state appears with $\rho_{yx}$ and $\sigma_{xy}$ showing full quantization and the vanishing $\rho_{xx}$ (i.e. 126Ω, 154Ω and 47 Ω for 3-5-3 (SH1), 3-4-3 and 3-6-3 respectively) in the parallel magnetization alignment regions. However, in antiparallel magnetization alignment regions, the axion insulator state shows up, where the $\rho_{xx}$ becomes very large (>40 $h/e^2$ for 3-5-3 (SH1) and 3-6-3, ~20 $h/e^2$ for 3-4-3), and $\sigma_{xy}$ shows a zero plateau, while $\rho_{yx}$ shows substantial fluctuations, as explained in the main text, the observed fluctuation is most likely due to the unavoidable electrical 'pickup' in $\rho_{yx}$ from the extremely large $\rho_{xx}$ with our imperfect Hall bar geometry.

For the 3-5-3(SH1) and the 3-4-3 sandwich heterostructures, the $V_g$ dependence of transport results is quite similar. For $V_g > V_g^0$, the zero $\sigma_{xy}$ plateau exists in antiparallel magnetization alignment regions, and the QAH state appears in parallel magnetization alignment regions. On the other hand, when $V_g < V_g^0$, in particular $V_g < 0$ V, the $\rho_{xx}$, $\rho_{yx}$ and $\sigma_{xy}$ are significantly altered, and their values become very small, indicating that the chemical potential of the sandwich heterostructure has crossed the bulk valence bands. When $V_g$ = -50 V, -100 V and -200 V, the two-step transition feature in $\rho_{yx}$ and $\sigma_{xy}$ persist in the 3-5-3 (SH1) sandwich heterostructure. However, it disappears in the 3-4-3 sandwich heterostructure, where the hysteresis loops become square-like, similar as these in uniformly doped TI films [3,4]. As we discussed in the main text, currently, the ferromagnetic order is modulated by the Ruderman-Kittel-Kasuya-Yosida (RKKY) interaction [4], so the ferromagnetic order should become stronger due to the introduction of substantial hole carriers in the system. When the interlayer magnetic coupling field $H_E$ is comparable to or larger than the difference of the $H_c$ of the top and bottom magnetic layers, the entire sandwich heterostructure is coupled. Therefore, the 3-4-3 sandwich heterostructure shows the uniform hysteresis loops when $V_g < 0$ V. The detailed comparisons of the $H_E$ in the three sandwich heterostructures are described in Section VII.

The 3-6-3 sandwich heterostructure shows an opposite $V_g$ dependent behavior. The QAH and axion insulator states are in the negative $V_g$s, and disappear when $V_g > 0$ V. Note that the $V_g^0$ of 3-6-3 sandwich heterostructure is ~ -16 V. It is very likely that in the 3-6-3 sandwich heterostructure, the chemical potential has crossed the bulk conduction bands when $V_g > 0$ V. When $V_g$ = +100 V and +200 V, the two-step transition feature in $\rho_{yx}$ and $\sigma_{xy}$ also appears, similar as that in 3-5-3 (SH1) sandwich heterostructure when $V_g < 0$ V.

## IV. Temperature dependence of magneto-transport results of the three sandwich heterostructures

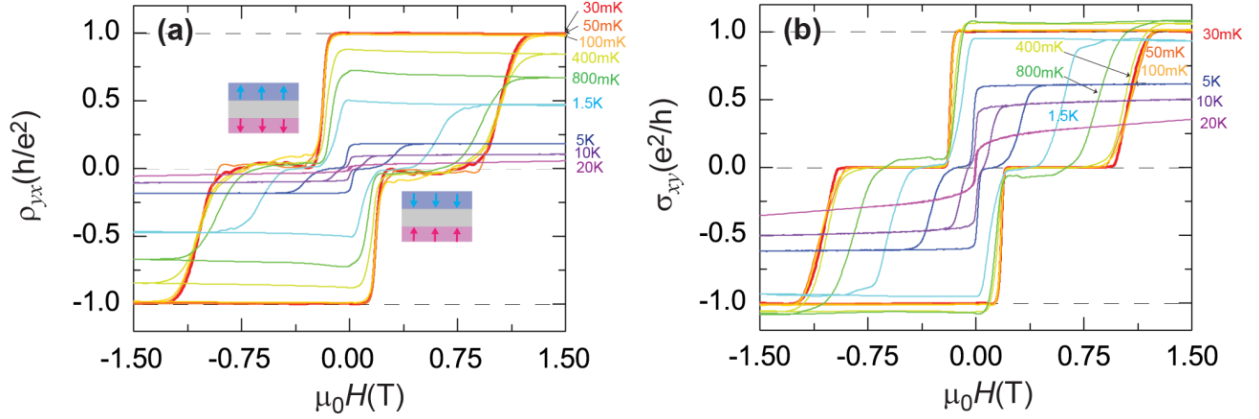

**Figure S9| Two-step magnetization reversals in 3-5-3 SH2.** (a, b) The $\mu_0 H$ dependence of $\rho_{yx}$ (a) and $\sigma_{xy}$ (b) at different temperatures. Both $\rho_{yx}$ and $\sigma_{xy}$ exhibit two-step magnetization transition for $T < 10$K.

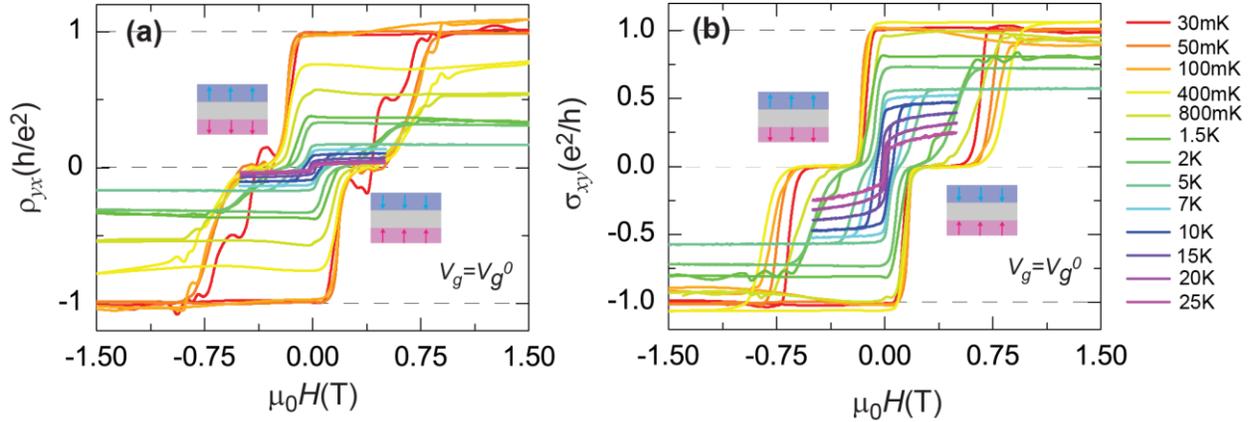

**Figure S10 | Two-step magnetization reversals in the 3-4-3 sandwich heterostructure.** (a, b) The $\mu_0 H$ dependence of $\rho_{yx}$ (a) and $\sigma_{xy}$ (b) at different temperatures. Both $\rho_{yx}$ and $\sigma_{xy}$ exhibit two-step magnetization transition for $T < 7$K. The 3-4-3 sandwich heterostructure used here and that shown in Fig. S7 are from the same wafer.

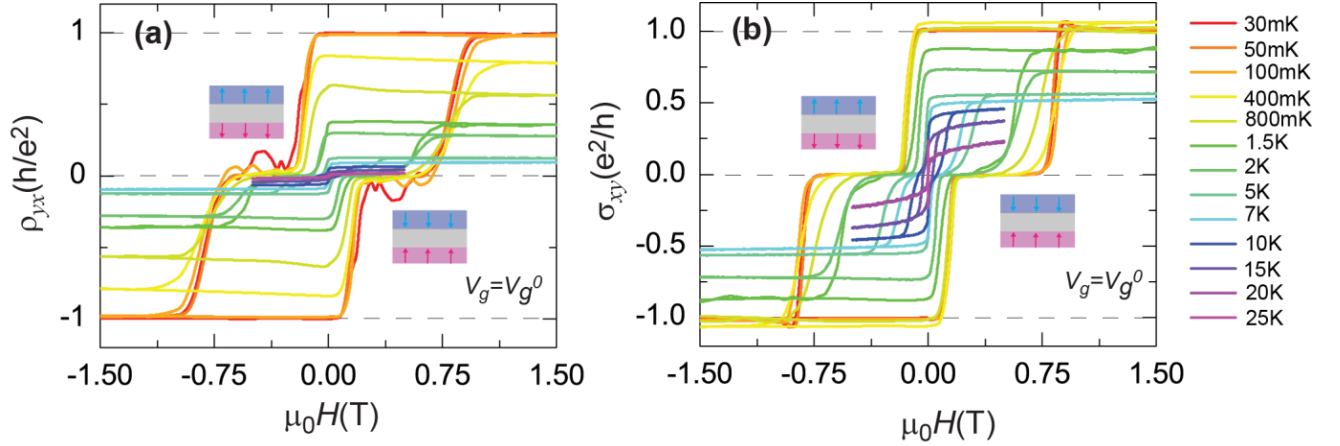

**Figure S11| Two-step magnetization reversals in the 3-6-3 sandwich heterostructure.** (a, b) The $\mu_0H$ dependence of $\rho_{yx}$ (a) and $\sigma_{xy}$ (b) at different temperatures. Both $\rho_{yx}$ and $\sigma_{xy}$ exhibit two-step magnetization transition for $T < 10$K. The 3-6-3 sandwich heterostructure used here and that shown in Fig. S8 are from the same wafer.

Figures S9-S11 show the $\mu_0H$ dependence of $\rho_{xx}$, $\rho_{yx}$ and $\sigma_{xy}$ at temperature range from 30 mK to 25 K for 3-5-3 (SH2) (Fig. 9), 3-4-3 (Fig. 10), and 3-6-3 (Fig. 11) sandwich heterostructures. The Hall traces in the three sandwich heterostructures show nonlinear behaviors at $T = 25$ K, indicating the Curie temperature ($T_C$) of the three sandwich heterostructures is ~ 25 K. The two-step transition feature disappears at $T = 10$ K for the 3-5-3(SH2) and the 3-6-3 sandwich heterostructures. For the 3-4-3 sandwich heterostructure, the two-step transition feature is much narrower and disappears at lower temperature, *i.e.* $T < 7$K, probably due to the thinner TI spacer (~4 QL) which induces larger $H_E$.

## V. Flow diagrams ($\sigma_{xy}$, $\sigma_{xx}$) of the three sandwich heterostructures

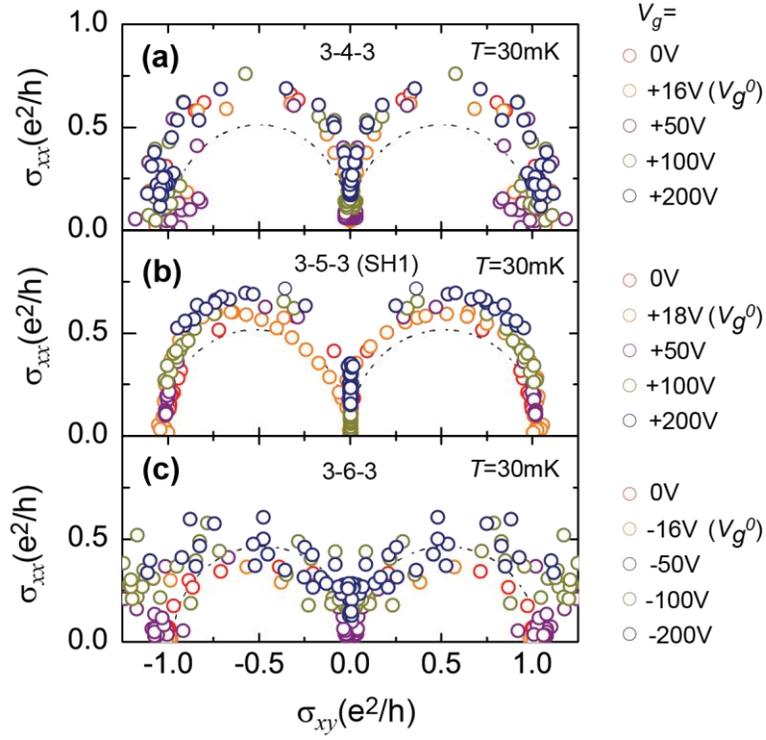

**Figure S12|** **Flow diagrams ($\sigma_{xy}$, $\sigma_{xx}$) of the three sandwich heterostructures at various $V_g$s and $T$ = 30 mK.** (a-c) Flow diagrams ($\sigma_{xy}$, $\sigma_{xx}$) of 3-4-3 (a), 3-5-3 (SH1) (b) and 3-6-3 (c) QAH sandwich heterostructures.

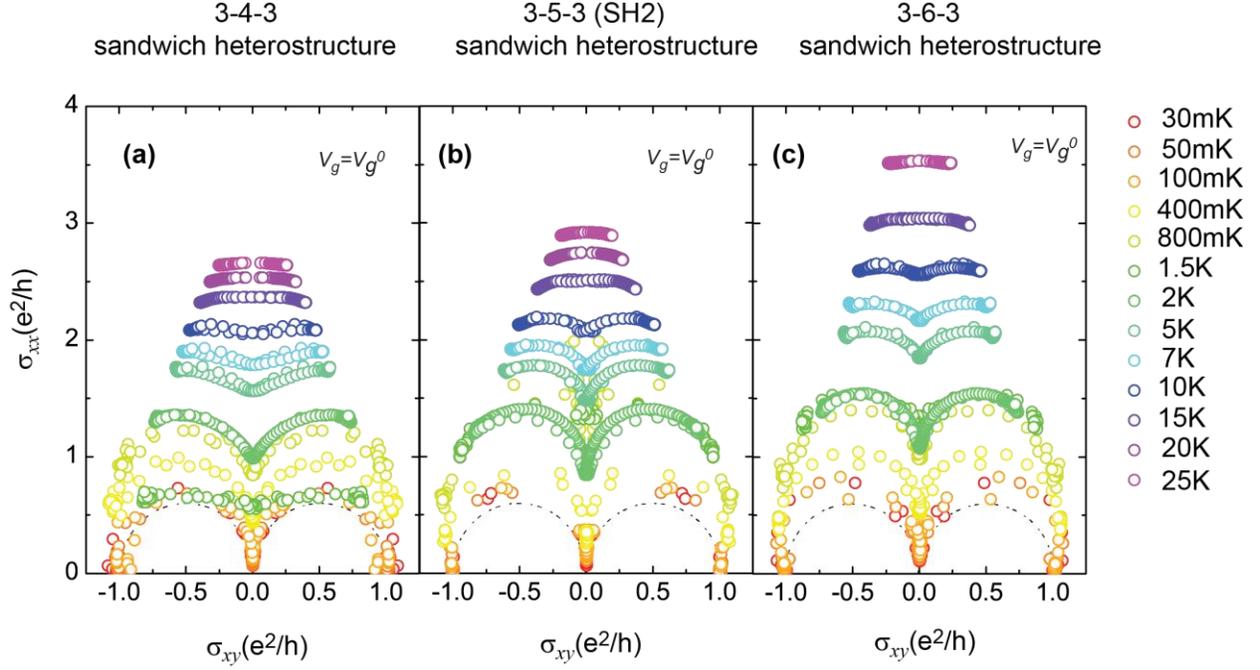

**Figure S13| Flow diagrams ($\sigma_{xy}$, $\sigma_{xx}$) of the three sandwich heterostructures at various $T$s and $V_g=V_g^0$.** (a-c) Flow diagrams ($\sigma_{xy}$, $\sigma_{xx}$) of 3-4-3 (a), 3-5-3 (SH2) (b) and 3-6-3 (c) QAH sandwich heterostructures.

Figure S12 shows the flow diagrams ($\sigma_{xy}$, $\sigma_{xx}$) of the three sandwich heterostructures (3-4-3, 3-5-3 (SH1) and 3-6-3) from $\mu_0H = 1.8$ T to $\mu_0H = -1.8$ T (extracted from the red curves in Figs. S6 to S8) at $T = 30$ mK and at different $V_g$s. The three sandwich heterostructures show very similar $V_g$ dependence of scaling behaviors. Two semicircles of 0.5 $e^2/h$ centered at ($\sigma_{xy}$, $\sigma_{xx}$) = ($\pm 0.5\ e^2/h$, 0) appear, as noted in the main text, the ($\sigma_{xy}$, $\sigma_{xx}$) = ($\pm e^2/h$, 0) and (0, 0) corresponds to the QAH insulator state (parallel magnetization alignment) and the axion insulator state (antiparallel magnetization alignment or zero $\sigma_{xy}$ plateau state), respectively.

Figure S13 shows the flow diagrams ($\sigma_{xy}$, $\sigma_{xx}$) of the three sandwich heterostructures (3-4-3, 3-5-3 (SH2) and 3-6-3) from $\mu_0H = 1.8$ T to $\mu_0H = -1.8$ T at $V_g = V_g^0$ and different temperatures.

The three sandwich heterostructures again show very similar $T$ dependence of scaling behaviors. The two-semicircle features, which correspond to the two-step transition feature in $\sigma_{xy}$ (Figs. S9 to S11), persist up to $T = 10$ K for the 3-5-3 (SH2) and the 3-6-3 sandwich heterostructures but only up to 7 K for the 3-4-3 sandwich heterostructure. For $T > 10$ K in the 3-5-3 (SH2) and the 3-6-3 sandwich heterostructures and for $T > 7$ K in the 3-4-3 sandwich heterostructure, the two-semicircle feature evolves to a single arc.

Note that the scaling of ($\sigma_{xy}$, $\sigma_{xx}$) of the three sandwich heterostructure resembles the global phase diagram of a 2D QAH sample with uniformly doping [5-7], but as we discussed in the main text, the underlying mechanisms for the zero $\sigma_{xy}$ plateau are distinctly different. In a uniformly doped 2D QAH sample, ($\sigma_{xy}$, $\sigma_{xx}$) = (0, 0) is the result of multi-domain formations induced insulating state during the magnetization reversal process [8]. However, in our QAH sandwich heterostructures, ($\sigma_{xy}$, $\sigma_{xx}$) = (0, 0) is the result of cancellation of top and bottom conduction in antiparallel magnetization alignment.

## VI. Two-terminal resistance measurements of the three sandwich heterostructures

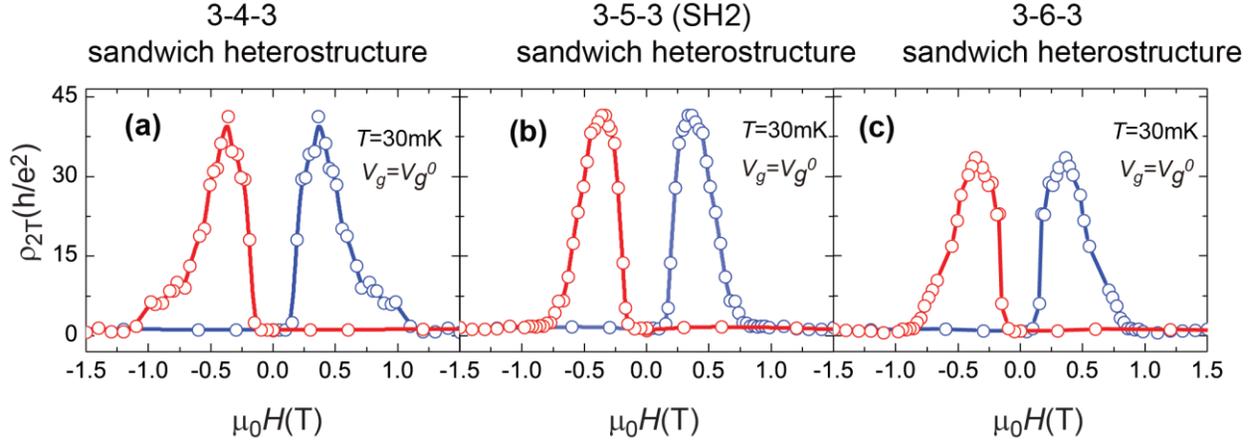

**Figure S14|** The $\mu_0 H$ dependence of two-terminal resistance $\rho_{2T}$ of the three sandwich heterostructure at $T = 30$ mK and $V_g = V_g^0$. (a-c) $\rho_{2T}$ of 3-4-3 (a), 3-5-3 (SH2) (b) and 3-6-3 (c) QAH sandwich heterostructures. The red and blue curves represent sweeping $\mu_0 H$ downward and upward, respectively.

. We also measured the $\mu_0 H$ dependence of the two-terminal resistance $\rho_{2T}$ of the three sandwich heterostructure at $T = 30$ mK and $V_g = V_g^0$, as shown in Fig. S14. The results for the three sandwich heterostructures show very similar $\mu_0 H$ dependence. In the parallel magnetization alignment (the QAH insulator state) regions, the $\rho_{2T}$ is close to the quantized value, $h/e^2$, consistent with $\rho_{2T}$ of uniformly doped QAH samples in the well-defined magnetization regions [9,10]. However, in the antiparallel magnetization (the axion insulator state) regions, the $\rho_{2T}$ becomes very large (~ 40 $h/e^2$), consistent with the insulating property of the axion insulator state, as discussed in the main text.

## VII. Minor loops of the three sandwich heterostructures

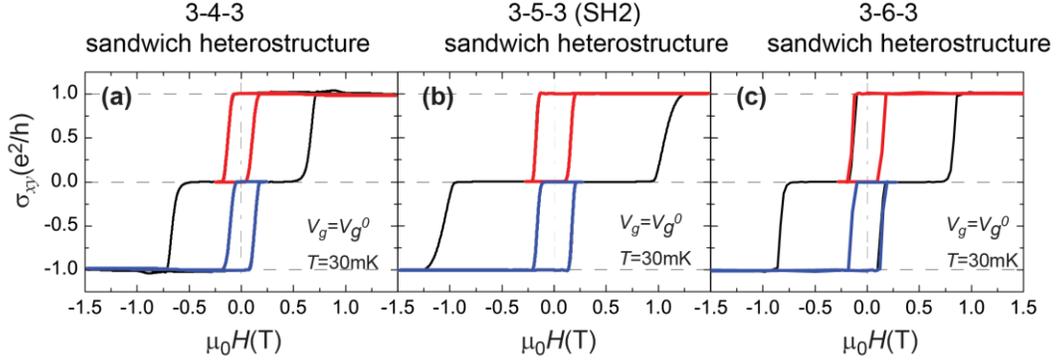

**Figure S15| Minor loops of $\sigma_{xy}$ of the three sandwich heterostructures at $T = 30$ mK and $V_g = V_g^0$.** (a-c) The minor loops of 3-4-3 (a), 3-5-3 (SH2) (b) and 3-6-3 (c) QAH sandwich heterostructures.

The minimum magnetic field, where the property of the hysteresis vanishes, is usually defined as the closure magnetic field. A "major loop" refers to a hysteresis loop measured with the starting and ending magnetic fields exceeding this closure magnetic field. When hysteresis is measured with the maximum magnetic field smaller than the closure magnetic field, the resulting loop is referred to as a "minor loop". Minor loop measurements usually provide additional information on the microscopic details of the magnetization reversal in ferromagnetic materials.

To further demonstrate the weak interlayer magnetic coupling effect, we measured the minor loops of the three QAH sandwich heterostructures at $T = 30$ mK and $V_g = V_g^0$, as shown in Fig. S15. In all the three QAH sandwich heterostructures, we found that the axion insulator state (the zero $\sigma_{xy}$ state) can be stabilized even at $\mu_0 H=0$. As discussed in the main text, the interlayer magnetic coupling field $H_E$ can be estimated by $H_E = \dfrac{\left|H_{c,\text{minor}}^L - H_{c,\text{minor}}^R\right|}{2}$, where $H_{c,\text{minor}}^L$ and $H_{c,\text{minor}}^R$ denote the left and right $H_c$ of minor loops [11,12]. The $H_E$s for the three sandwich heterostructure at $T = 30$ mK and $V_g = V_g^0$ are ~135 Oe (3-4-3), ~115 Oe (3-5-3 (SH2)) and ~65 Oe (3-6-3), respectively. It is obvious that, by increasing the thickness of the undoped TI spacer, $H_E$ is reduced. In each sandwich heterostructure, the $H_E$ is one order of magnitude smaller than the difference between $H_{c1}$ and $H_{c2}$, in agreement with the existence of the antiparallel magnetization alignment at $T = 30$ mK.

When the thickness of the spacer layer is further increased, the interlayer magnetic coupling effect is weakened further, but the additional bulk dissipative channels or the helical side surface states will possibly be introduced. Therefore, we expect it will be more challenging to realize the QAH state in parallel magnetization alignment and the axion insulator state in antiparallel magnetization alignment in the sandwich heterostructure with the total thickness > 12 QL.

## VIII. MFM images of the 3-5-3 SH2

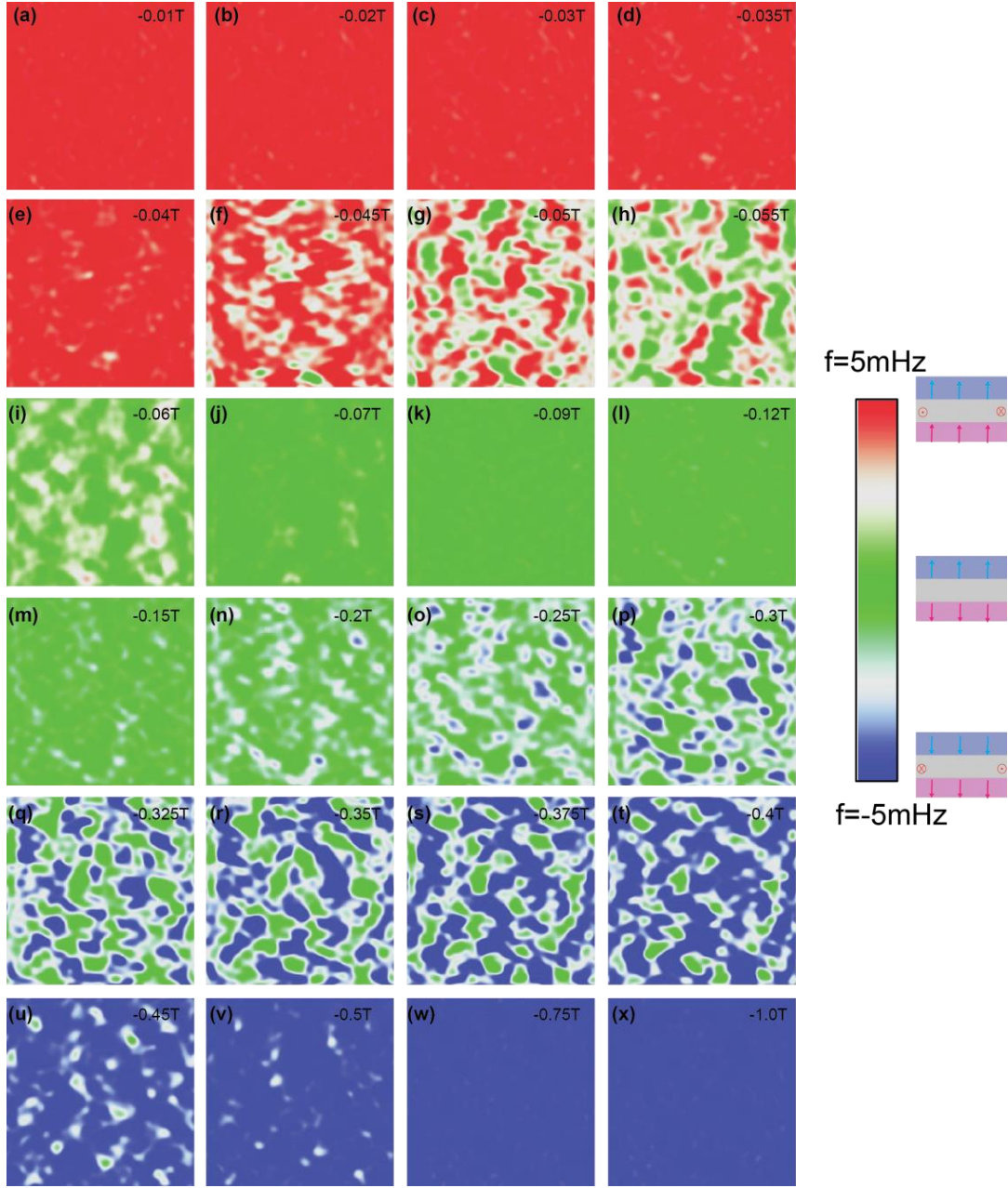

**Figure S16| Two-step magnetization reversals in the 3-5-3 SH2.** (a-x) The MFM images at various $\mu_0H$s. Red and blue represent upward and downward parallel magnetization alignment regions, green represents antiparallel magnetization alignment regions. All the MFM images are taken at $V_g = 0$ V and $T = 5.3$ K. The MFM images show two-step magnetization transitions from

-0.01 T to -1 T, demonstrating the antiparallel magnetization alignment of top and bottom magnetic layers between two $H_c$s at $T$ = 5.3 K. All the MFM images and the Hall traces in Fig. 3a of the main text are measured simultaneously. The magnetic domain contrast ($\delta f$) at each $\mu_0 H$ in Fig. 3b of the main text is estimated by the root-mean-square (RMS) value of the corresponding MFM image in (a-x).

Figure S16 shows the MFM images of the 3-5-3 SH2 at various $\mu_0 H$s at $T$ = 5.3 K and $V_g$ = 0 V. As discussed in the main text, when the $\mu_0 H$ is swept from -0.01 T to -1.0 T, the MFM images show clear two-step magnetization transitions, one is around $\mu_0 H$ = -0.055 T and the other is around $\mu_0 H$ = -0.375 T. The former corresponds to the magnetization reversal of the bottom 'softer' Cr-doped TI layer. The latter corresponds to the magnetization reversal of the top 'harder' V-doped TI layer. When $\mu_0 H$ is swept between the two $H_c$s *(i.e.* -0.375 T < $\mu_0 H$ < -0.055 T at *T*=5.3 K), the antiparallel magnetization alignment regions (green color regions) are dominated in the sandwich heterostructure. This antiparallel magnetization alignment state exists in 3-5-3 SH2 at other *T*s and $V_g$s if the relative $H_E$ value $\alpha$ is smaller than 0.028, which is the $\alpha$ value for this sandwich heterostructure at $T$ = 5.3 K and $V_g$ = 0 V.

# References

[1] J. S. Zhang, C. Z. Chang, Z. C. Zhang, J. Wen, X. Feng, K. Li, M. H. Liu, K. He, L. L. Wang, X. Chen, Q. K. Xue, X. C. Ma, and Y. Y. Wang, *Nat. Commun.* **2**, 574 (2011).

[2] C. Z. Chang, J. S. Zhang, X. Feng, J. Shen, Z. C. Zhang, M. H. Guo, K. Li, Y. B. Ou, P. Wei, L. L. Wang, Z. Q. Ji, Y. Feng, S. H. Ji, X. Chen, J. F. Jia, X. Dai, Z. Fang, S. C. Zhang, K. He, Y. Y. Wang, L. Lu, X. C. Ma, and Q. K. Xue, *Science* **340**, 167 (2013).

[3] C. Z. Chang, J. S. Zhang, M. H. Liu, Z. C. Zhang, X. Feng, K. Li, L. L. Wang, X. Chen, X. Dai, Z. Fang, X. L. Qi, S. C. Zhang, Y. Y. Wang, K. He, X. C. Ma, and Q. K. Xue, *Adv. Mater.* **25**, 1065 (2013).

[4] Z. C. Zhang, X. Feng, M. H. Guo, K. Li, J. S. Zhang, Y. B. Ou, Y. Feng, L. L. Wang, X. Chen, K. He, X. C. Ma, Q. K. Xue, and Y. Y. Wang, *Nat. Commun.* **5**, 4915 (2014).

[5] X. F. Kou, L. Pan, J. Wang, Y. B. Fan, E. S. Choi, W. L. Lee, T. X. Nie, K. Murata, Q. M. Shao, S. C. Zhang, and K. L. Wang, *Nat. Commun.* **6**, 8474 (2015).

[6] S. Grauer, K. M. Fijalkowski, S. Schreyeck, M. Winnerlein, K. Brunner, R. Thomale, C. Gould, and L. W. Molenkamp, *Phys. Rev. Lett.* **118**, 246801 (2017).

[7] Y. Feng, X. Feng, Y. B. Ou, J. Wang, C. Liu, L. G. Zhang, D. Y. Zhao, G. Y. Jiang, S. C. Zhang, K. He, X. C. Ma, Q. K. Xue, and Y. Y. Wang, *Phys. Rev. Lett.* **115**, 126801 (2015).

[8] J. Wang, B. Lian, and S. C. Zhang, *Phys. Rev. B* **89**, 085106 (2014).

[9] A. J. Bestwick, E. J. Fox, X. F. Kou, L. Pan, K. L. Wang, and D. Goldhaber-Gordon, *Phys. Rev. Lett.* **114**, 187201 (2015).

[10] C. Z. Chang, W. W. Zhao, D. Y. Kim, P. Wei, J. K. Jain, C. X. Liu, M. H. W. Chan, and J. S. Moodera, *Phys. Rev. Lett.* **115**, 057206 (2015).

[11] P. Walser, M. Hunziker, T. Speck, and M. Landolt, *Phys. Rev. B* **60**, 4082 (1999).

[12] J. Faure-Vincent, C. Tiusan, C. Bellouard, E. Popova, M. Hehn, F. Montaigne, and A. Schuhl, *Phys. Rev. Lett.* **89**, 107206 (2002).